\documentclass[aps,showpacs,superscriptaddress,twocolumn,floatfix]{revtex4}

\usepackage{epsfig}
\usepackage{latexsym}   
\usepackage{amssymb} 	
\usepackage{amsfonts} 	
\usepackage{amsthm} 	
\usepackage{amsmath} 	

\newcommand{{\vp}}{{\vec p}} 
\newcommand{{\vq}}{{\vec q}}

\newcommand{\beq}{\begin{equation}}
\newcommand{\eeq}{\end{equation}} 

\newcommand{\lton}{\mathrel{\lower.9ex \hbox{$\stackrel{\displaystyle<}{\sim}$}}} 
\newcommand{\ee}{\end{equation}}
\newcommand{\ben}{\begin{enumerate}} 
\newcommand{\een}{\end{enumerate}}
\newcommand{\bit}{\begin{itemize}} 
\newcommand{\eit}{\end{itemize}}
\newcommand{\bc}{\begin{center}} 
\newcommand{\ec}{\end{center}}
\newcommand{\bea}{\begin{eqnarray}} 
\newcommand{\eea}{\end{eqnarray}}
\newcommand{\beqar}{\begin{eqnarray}} 
\newcommand{\eeqar}[1]{\label{#1} \end{eqnarray}}

\newcommand{\eq}[1]{Eq.~(\ref{#1})}
\newcommand{\fig}[1]{Fig.~\ref{#1}}

\newcommand{\raa}{R_{AA}}
\newcommand{\av}[1]{ \langle #1 \rangle}

\newcommand{\bb}[1]{ {\mathbf #1 }}

\begin{document}

\renewcommand{\topfraction}{0.85}
\renewcommand{\textfraction}{0.1}
\renewcommand{\floatpagefraction}{0.75}

\title{Up to and beyond ninth order in opacity: Radiative energy loss with GLV}

\date{\today  \hspace{1ex}}

\author{Simon Wicks}
\email{simonw@phys.columbia.edu}
\affiliation{Dept. Physics, Columbia University, 538 W 120-th Street,\\ New York,
       NY 10027, USA}

\begin{abstract}
A new examination of the GLV all-orders opacity result for radiative energy loss is presented. The opacity expansion is shown to be a Dyson expansion of a Schrodinger-like (or diffusion) equation, a form also found in BDMPS-Z-ASW, AMY and Higher Twist formalisms of radiative energy loss. A new, efficient numerical evaluation of the GLV all-orders result gives results up to and beyond ninth order in opacity. This presents a solution to this Schrodinger-like equation to almost arbitrary accuracy for lengths and densities applicable at RHIC and LHC. The first order is found to be most important, although higher orders implemented in a geometry integration will be needed for quantitative jet tomography.
\end{abstract}

\maketitle

\section{Introduction}
Jet quenching theory was buoyed by early success at RHIC. This was based on the application of calculations based on perturbative quantum chromodynamics (pQCD) to high momentum particles (jets) traversing a quark-gluon plasma (QGP) at mid-rapidity produced in Au-Au collisions at $\sqrt{s} = 62-200$AGeV~\cite{Gyulassy:1993hr,Gyulassy:2003mc,Kovner:2003zj,Jacobs:2004qv}. Reasonable choices of input parameters could predict or explain high momentum pion and eta $\raa(p_T)$ data out to $p_T \approx 20$GeV~\cite{Adler:2003qi,Isobe:2005mh,Shimomura:2005en}.

These calculations were based on radiative energy loss. Coherence effects due to the soft emitted gluon make radiative calculations a difficult business. There are currently four different calculations on the market for radiative energy loss in a QGP: (D)GLV~\cite{Gyulassy:1999zd,Gyulassy:2000fs,Gyulassy:2000er,Gyulassy:2001nm,Djordjevic:2003zk}, BDMPS(-Z-ASW)~\cite{Baier:1996kr,Baier:1996sk,Baier:1998yf,Zakharov:1996fv,Zakharov:1997uu,Baier:1998kq,Wiedemann:2000ez,Wiedemann:2000za,Wiedemann:2000tf,Salgado:2003gb,Armesto:2005iq}, Higher Twist~\cite{Wang:2001ifa,Guo:2000nz,Osborne:2002st,Zhang:2003yn}, AMY~\cite{Arnold:2002ja,Jeon:2003gi,Turbide:2005fk}. In spite of the similar roots of all the calculations, little agreement between groups can be found, in results or even use of language. The long term disagreement between four similar schemes presents significant difficulties for the interpretation of experimental data at RHIC and the upcoming LHC. In the PHENIX study \cite{Adare:2008cg}, this is represented as an approximate factor ten uncertainty in the extracted medium parameter. Clearly we have not yet achieved quantitative jet tomography.

The strongly coupled techniques of AdS/CFT have recently surged onto the jet quenching scene~\cite{Gubser:2006bz,Herzog:2006gh,Liu:2006ug,CasalderreySolana:2006rq,Horowitz:2007su}. Many of these implicitly reject the notion of radiative energy loss, the emission of relatively soft quasi-particles. The heavy quark puzzle, the unexpectedly~\cite{Djordjevic:2005db} strong quenching of heavy quarks seen through the low $\raa(p_T)$ of single non-photonic electrons~\cite{Adler:2005xv,Akiba:2005bs,Abelev:2006db,Adare:2006hc,Bielcik:2005wu,Dong:2005nm,Adare:2006nq}, has also motivated broader investigation of energy loss mechanisms. Possible `solutions' include a re-examination of collisional energy loss~\cite{Wicks:2005gt,Adil:2006ei,Wicks:2007am,Wicks:2007zz}, heavy meson dissociation mechanisms~\cite{Adil:2006ra}, resonances in the medium~\cite{vanHees:2005wb,vanHees:2004gq,Rapp:2008qc}, or the above-mentioned AdS/CFT techniques. A consistent theory of radiative energy loss is important to indicate whether further mechanisms are necessary.

This paper begins to bridge the gap between radiative energy loss formalisms. This begins in section \ref{sec:closerglv} by emphasizing the common foundations of all the calculations of gluon radiation. The GLV result is expressed as an opacity expansion. This is shown to be a Dyson expansion of a Schrodinger-like (or diffusion) equation. This Schrodinger-like equation is not confined to thick plasmas or uncorrelated medium assumptions, and has been used in other radiative energy loss formalisms.

This in itself is not new. But the use of the opacity expansion is twofold: in the derivation of the Schrodinger-like equation (including the explicit demonstration of its color triviality), and as a means to its numerical solution. The GLV result was expressed in closed form to all orders in opacity. A new numerical evaluation of this result is presented in section \ref{sec:rad-allorders}, which is able to find results up to and beyond ninth order in opacity. 

The numerical results presented here use a Gyulassy-Wang (GW) static color center model of the medium. Possible improvements to this model have been suggested Djordjevic and Heinz's dynamical medium calculations~\cite{Djordjevic:2007at,Djordjevic:2008iz}. The formalism and numerical solution presented here are not limited to this GW model, and future studies can explore further possibilities. For example, Vitev has given all-orders expressions for initial state, final state and infinite time radiation~\cite{Vitev:2007ve}.

In this way, for lengths and densities applicable at RHIC and LHC, all the significant terms in the series are found. The interference with the creation radiation is crucial to an accurate evaluation of the radiation spectrum. Other numerical solutions to similar Schrodinger-like equations have been presented by other groups, and in future these solutions can be compared to the method of solution found here. There is no need to sit still with a fixed thin or thick plasma approximation. Once this is solved for a fixed length and density, the incorporation of these results into geometry integrations can mark the beginning of the era of quantitative jet tomography.

\section{The GLV opacity expansion}
\label{sec:closerglv}
In GLV II~\cite{Gyulassy:2000er} a reaction operator was derived, used to iterate from order to order in opacity and to prove the `color triviality' of this procedure, and to derive a closed form expression for an all orders in opacity result. Here, we look again at the operator recursion in opacity, show its equivalence to a Schrodinger-like equation (for both thin and thick plasmas, correlated and uncorrelated media) similar to those used in other radiative energy loss formalisms, and discuss the significance of this result.

\subsection{The Schrodinger-like equation}
\label{sec:formulaglvii}
The GLV recursion relation is derived in GLV II~\cite{Gyulassy:2000er}. This section continues on from that derivation. It uses eikonal kinematics for both the jet and the emitted gluon
\begin{equation}
 k_\perp \ll \omega \mbox{ , } E \gg q,q^0
\end{equation}
and a small $x$ approximation - the energy of the emitted gluon is much less than the energy of the jet. The longitudinal momentum of the gluon at different stages is written as $\omega_i$,
\begin{equation}
 \omega_0 = \frac{\bb{k}^2}{2 x E} \mbox{ ,  } \omega_i = \frac{(\bb{k}-\bb{q}_i)^2}{2 x E}
\end{equation}
This assumes that the energy transfer with the medium is zero, or at least negligible compared to the other terms.

The GLV result expresses the relation between successive orders in opacity:
\begin{align}
\label{eqn:pnk}
 P_n({\mathbf k}) &= C_A \left[ P_{n-1}(\bb{k}-\bb{q}_n) - P_{n-1}(\bb{k}) \right] \nonumber \\     
	      &-2 C_A \bb{B}_n .  \left( \mbox{Re} e^{-i\omega_n z_n}  e^{i \bb{q}_n . \hat{ \bb{ b } } }  \bb{I}_{n-1}  \right) \nonumber \\
	      &+ \delta_{n,1} C_A C_R | \bb{B}_1 |^2
\end{align}
where $P_n(\bb{k})$ is, up to multiplicative constants, the emission probability of a gluon of momentum $\bb{k}$ (where $\bb{k}$ is momentum transverse to the direction of the jet).

The first line is identified as a classical cascade. Once the emitted gluon has decohered from the jet, it collides with the medium and diffuses in 2 dimensional $\bb{k}$ (=$\vec{k_\perp}$) space. This does not change the total number, it integrates to zero. 

The second line is identified as a quantum cascade. It acts as a source term into the classical cascade. Each possible collision evolves the amplitude and its complex conjugate, giving evolution in $\bb{I}_n$. This recursion is
\begin{align}
\label{eqn:in}
 \bb{I}_n &= C_A \left( e^{i(\omega_0 - \omega_n)z_n} e^{i \bb{q}_n.\hat{ \bb{ b } }} - 1 \right) \bb{I}_{n-1} \nonumber \\
 	   & \; - \delta_{n-1} C_A C_R \bb{B}_1 e^{i \omega_0 z_1}
\end{align}
The first line iterates from collision to collision, the second only applies to the first collision. The source from the second collision is smaller than (LPM suppression) or equal to (the incoherent limit) the first collision. 

The last line in \eq{eqn:pnk} is a source from the first collision. We rewrite \eq{eqn:pnk} as a classical cascade with a source term from a quantum process,
\begin{align}
 P_n(\bb{k}) &= P_{n,\mbox{classical}}(\bb{k}) \nonumber \\
  &\; + S_{n,\mbox{quantum}}(\bb{k}) + \delta_{n,1} X_{1,\mbox{quantum}}(\bb{k})
\end{align}
$S_{n,\mbox{quantum}}$ will be written as $S_n$. The iterations need initial conditions, ie $P_0$ and $\bb{I}_0$. These differ depending on whether the incoming jet is on-shell, ie created at $t=-\infty$, or created at $t=t_0$. For the on-shell jet,
\begin{align}
 P_0(\bb{k}) = 0 \mbox{ ,  } \bb{I}_0 = \bb{0}
\end{align}
For the finite time case,
\begin{align}
 P_0(\bb{k}) = C_R \bb{H}^2 \mbox{ ,  } \bb{I}_0 = -C_R \bb{H} e^{i \omega_0 z_0}
\end{align}
Note how a $C_A$ is always associated with an interaction with the medium, the $C_R$ is associated with the emission vertex.

\subsection{Rearranging the formulae}
The shift operator appears in the iteration for $\bb{I}_n$, \eq{eqn:in}. This form will appear frequently:
\begin{equation}
 \hat{O}_i = C_A ( e^{i (\omega_0 - \omega_i) z_i} e^{i \bb{q_i}.\hat{ \bb{ b } }} - 1 )
\end{equation}
The first term corresponds to the phase shift from the (soft) gluon colliding with the medium. The second term corresponds to no phase shift from the (hard) jet colliding with the medium. At this stage, the operator $\hat{O_i}$ is a function of the position $z_i$, the momentum transfer $\bb{q}_i$ and the gluon momentum $\bb{k}$. Averaging over $\bb{q}_i$ and $z_i$ will come later. The shift operator is remarkably simple, hiding its complex derivation through the evaluation of direct and virtual QCD diagrams. This form is specific for the eikonal approximation, with no (or negligible) energy exchange with the medium. Note that the shift operator at $z_i$ is only dependent on the momentum transfer $\bb{q}_i$, and independent of the rest.

First, we want to simplify the beginning and end of the series giving $S_{n}$. $\bb{B}_i$ can be rewritten with use of the shift operator on a hard vertex:
\begin{equation}
 C_A \bb{B}_i e^{i \omega_0 z_i} = - \hat{O}_i \bb{H} e^{i \omega_0 z_i}
\end{equation}
In this way, we can express the first and last collisions in terms of the shift operator
\begin{align}
  - \hat{O}_1 \bb{J}_i  &= \bb{I}_1\\
  \bb{J}_f \hat{O}_n  &= C_A \bb{B}_n e^{-i \omega_n z_n} e^{i \bb{q}_n.\hat{ \bb{ b } }} 
\end{align}
where
\begin{align}
 \bb{J}_f(z_n) &= - C_R \bb{H} e^{-i \omega_0 z_n} \\
 \bb{J}_i(z_0,z_1) &= - C_R \bb{H} e^{-i \omega_0 z_1} \nonumber \\
  & \; \mbox{ for a jet created at t=$-\infty$} \\
 \bb{J}_i(z_0,z_1) &= - C_R \bb{H} ( e^{-i \omega_0 z_1} - e^{-i \omega_0 z_0} ) \nonumber \\
  & \; \mbox{ for a jet created at t=0}
\end{align}
Now we have the quantum cascade contribution to the gluon spectrum in a simpler form:
\begin{equation}
 S_n(\bb{k}) = -2 \bb{J}_f(z_n) \, \hat{O}_n \hat{O}_{n-1} \ldots \hat{O}_1 \, \bb{J}_i(z_0,z_1)
\end{equation}

This will be averaged over $q_i$ and $z_i$. First, we look at $q_i$, and define
\begin{align}
 \av{\hat{O}_i} &= \av{\hat{O}(z_i)}_{\bb{q}_i} \nonumber \\
  &= \int d^2q_i \frac{\bar{V}^2(z_i,q_i)}{\lambda(z_i)} (e^{i (\omega_0 - \omega_i) z_i} e^{i \bb{q}_i.\hat{ \bb{ b } }} - 1) \nonumber \\
  &= \int d^2q_i \frac{\bar{V}^2(z_i,q_i)-\delta^2(q_i)}{\lambda(z_i)} e^{i (\omega_0 - \omega_i) z_i} e^{i \bb{q}_i.\hat{ \bb{ b } }}
\end{align}
This is the momentum averaged shift operator. It is local in $z$. The potential has been normalized to give $\bar{V}$, with the mean free path beneath it (the $C_A$ has been absorbed into this $\lambda(z)$). Remembering that $\av{\ldots}$ represents averaging over momentum transfers (not the $z_i$s yet), the expression for $S_n$ now becomes
\begin{equation}
 \av{S_n(\bb{k})} = -2 \bb{J}_f(z_n) \av{\hat{O}_n} \av{\hat{O}_{n-1}} \ldots \av{\hat{O}_1} \bb{J}_i(z_0,z_1)
\end{equation}

We now integrate over the $z_i$s.
\begin{align}
 \av{S_n(\bb{k})}_{\bb{q}_i,z_i} &= -2 \int_{z_0}^{z_f} dz_1 \int_{z_1}^{z_f} dz_2 \ldots \int_{z_n}^{z_f} dz_n  \times \nonumber \\
  & \; \bb{J}_f(z_n) \av{\hat{O}_n} \av{\hat{O}_{n-1}} \ldots \av{\hat{O}_1} \bb{J}_i(z_0,z_1)
\end{align}
All the $z_i$ integrations pair up with their $\hat{O}_i$, except the first and last, $z_1$ and $z_n$ - as the currents $\bb{J}_i(z_0,z_1)$ and $\bb{J}_f(z_n)$ introduce additional $z$ dependence. Now the aim is to pull the $dz_n$ integration out to the front. Using
\begin{equation}
 \int_{z_{n-2}}^{z_f} dz_{n-1} \int_{z_{n-2}}^{z_f} dz_n = \int_{z_{n-2}}^{z_f} dz_n \int_{z_{n-2}}^{z_n} dz_{n-1}
\end{equation}
this gives
\begin{align}
 \av{S_n(\bb{k})}_{q_i,z_i} = -2 &\int_{z_0}^{z_f} dz_n \bb{J}_f(z_n) \av{\hat{O}_n} \times \nonumber \\
&\int_{z_0}^{z_n} dz_1 \int_{z_1}^{z_n} dz_2 \ldots \int_{z_{n-2}}^{z_n} dz_{n-1} \times \nonumber \\
  & \;\; \av{\hat{O}_{n-1}} \ldots \av{\hat{O}_1} \bb{J}_i(z_0,z_1)
\end{align}
Defining $\boldsymbol{\psi}_n$,
\begin{align}
 \boldsymbol{\psi}_n(z_0,z_n) = \int_{z_0}^{z_n} &dz_1 \int_{z_1}^{z_n} dz_2 \ldots \int_{z_{n-2}}^{z_n} dz_{n-1} \times \nonumber \\ & \; \av{\hat{O}_{n-1}} \ldots \av{\hat{O}_1} \bb{J}_i(z_0,z_1)
\end{align}
gives
\begin{equation}
 \av{S_n(\bb{k})}_{q_i,z_i} = -2 \int_{z_0}^{z_f} dz_n \bb{J}_f(z_n) \av{\hat{O}_n} \boldsymbol{\psi}_n(z_0,z_n)
\end{equation}

\subsection{Summing over opacities}
The result will be summed over all possible opacities, from $n=1\rightarrow \infty$. The relevant opacity weights come out from the calculation, as the $\lambda_i$s have already been included. The result is
\begin{equation}
 \sum_{n=1}^{\infty} \av{S_n(\bb{k})}_{q_i,z_i} = -2 \sum_{n=1}^{\infty} \int_{z_0}^{z_f} dz_n \bb{J}_f(z_n) \av{\hat{O}_n} \boldsymbol{\psi}_n(z_0,z_n)
\end{equation}
The $z_n$ variable as a dummy variable can be replaced by $z_e$, and pulled out of the sum, giving
\begin{equation}
 \sum_{n=1}^{\infty} \av{S_n(\bb{k})}_{q_i,z_i} =  -2 \int_{z_0}^{z_f} dz_e \bb{J}_f(z_e) \av{\hat{O}_e} \sum_{n=1}^{\infty} \boldsymbol{\psi}_n(z_0,z_e)
\end{equation}
Defining the full wave function
\begin{equation}
 \boldsymbol{\psi}(z_0, z_e) = \sum_{n=1}^\infty \boldsymbol{\psi}_n(z_0,z_e)
\end{equation}
gives the equation for $\boldsymbol{\psi}$ as
\begin{align}
 \boldsymbol{\psi}(z_0,z_e) = \sum_{n=1}^\infty &\int_{z_0}^{z_e} dz_1 \int_{z_1}^{z_e} dz_2 \ldots \int_{z_{n-2}}^{z_e} dz_{n-1} \times \nonumber \\
  & \; \av{\hat{O}_{n-1}} \ldots \av{\hat{O}_1} \bb{J}_i(z_0,z_1)
\end{align}
Written out as a series, this becomes
\begin{align}
 \boldsymbol{\psi}&(z_0,z_e) =  \int_{z_0}^{z_e} dz_1 \ldots \av{\hat{O}_1} \bb{J}_i(z_0,z_1) \nonumber \\
	&+ \int_{z_0}^{z_e} dz_1 \int_{z_1}^{z_e} dz_2 \av{\hat{O}_2} \av{\hat{O}_1} \bb{J}_i(z_0,z_1) \nonumber \\
	&+ \int_{z_0}^{z_e} dz_1 \int_{z_1}^{z_e} dz_2 \int_{z_2}^{z_e} dz_3  \av{\hat{O}_3} \av{\hat{O}_2} \av{\hat{O}_1} \bb{J}_i(z_0,z_1)  \nonumber \\
 	&+ \ldots
\end{align}
This is familiar from scattering theory, reminiscent of a Dyson expansion. 
\begin{equation}
 \boldsymbol{\psi} (z_0,z_e) = T \left[ \exp \left( \int_{z_0}^{z_e} dz \av{\hat{O}}(z) \right) \right]
\end{equation}
Hence, $\boldsymbol{\psi}(z_0,z_e)$ satisfies a Schrodinger-like equation
\begin{equation}
 \frac{\partial}{\partial z} \boldsymbol{\psi}(z,\bb{k}) = \av{\hat{O}}(z,\bb{k}) \boldsymbol{\psi}(z,\bb{k})
\end{equation}

\subsection{The Hamiltonian in impact parameter space}
In q-space, the form of the operator is unfamiliar, but a transformation into impact parameter space makes the Schrodinger analogy clearer.

A change of phase to
\begin{equation}
 \boldsymbol{\psi}(z,\bb{k}) = e^{i \omega_0 z} i \boldsymbol{\phi}(z,\bb{k})
\end{equation}
gives
\begin{equation}
 e^{i \bb{q}.\hat{ \bb{ b } }} \boldsymbol{\psi}(z,\bb{k}) = i e^{i \omega z} \boldsymbol{\phi}(z,\bb{k}-\bb{q})
\end{equation}
and a Fourier transform
\begin{align}
 \boldsymbol{\tilde{\phi}} (z,\bb{b}) &= \int d^2k e^{-i \bb{b}.\bb{k}} \boldsymbol{\phi}(z,\bb{k}) \\
 \tilde{V}(z,\bb{b}) &= \int d^2q e^{-i \bb{b}.\bb{q}} \bar{V}^2(z,\bb{q})
\end{align}
gives the equation
\begin{equation}
 i \frac{\partial \boldsymbol{\tilde{\phi}}}{\partial z} = \left( \frac{-1}{2 x E } \nabla_b^2 + i \frac{1}{\lambda(z)} ( \tilde{V}(z,\bb{b}) - 1 ) \right) \boldsymbol{\tilde{\phi}}
\end{equation}
This is a Schrodinger-like or diffusion equation in two dimensions with an imaginary (ie absorbing) potential. The Hamiltonian is
\begin{equation}
 \hat{H} = \frac{-1}{2 x E } \nabla_b^2 + i \frac{1}{\lambda(z)} ( \tilde{V}(z,\bb{b}) - 1 )
\end{equation}
This was expressed by Zakharov~\cite{Zakharov:2000iz} as
\begin{equation}
 \hat{H}(z) = - \frac{1}{2M(x)} \frac{\partial^2}{\partial \boldsymbol{\rho}^2} - i \frac{n(z) \sigma_3(\rho,z)}{2}
\end{equation}
where $\boldsymbol{\rho} = \bb{b}$, and $\sigma_3$ is the `cross section for interaction of the $\bar{q}qg$ system with a scattering center'. This can be used in path integral form, with the Lagrangian
\begin{equation}
 {\mathcal L} = \frac{1}{2} M(x) \dot{\boldsymbol{\rho}}^2 - \frac{1}{2}i  n(z)\sigma_3(\boldsymbol{\rho},z)
\end{equation}
Wiedemann describes this eikonal expression as the non-Abelian Furry approximation~\cite{Wiedemann:2000ez}. The $\dot{\rho}^2$ kinetic term comes from the eikonal $\omega_0 = k^2 / 2 x E$ - terms beyond the eikonal will spoil the simple diffusion equation. The AMY result~\cite{Jeon:2003gi} is expressed as an integral equation, which can be recast as a Schrodinger-like differential equation. The Higher Twist expansion has been `resummed' recently by Majumder, in application to photon Bremsstrahlung~\cite{Majumder:2007ne}, to give a similar diffusion equation.

\subsection{Interpreting the Schrodinger equation}
For the intuitive interpretation of the Schrodinger equation, it must be remembered that it is not simply a Schrodinger equation for the propagation of the jet. Zakharov goes into much detail of interpretation of this Schrodinger-like equation. From the GLV recursion, it is clear that the evolution is not just in the amplitude, but also in its conjugate. This is far simpler due to the color triviality produced - as shown in GLV, the reaction operator is color trivial, and hence the evolution considered simultaneously in the amplitude and its conjugate does not have complicated color dependence.

An opacity expansion is a perturbative-like Dyson expansion of the full result. Hence, the GLV opacity expansion recursion is equivalent in form to other approaches that started from the Schrodinger-like approach, ie BDMPS and Zakharov. The Schrodinger form does not assume a thick plasma, only the original approximated solutions of the equation do so - in a similar way to how a Schrodinger equation in non-relativistic quantum mechanics can describe one, two, or an arbitrary number of interactions. BDMPS derives the form of the equation, starting with a large $N_c$ result and then giving the general result~\cite{Baier:1996kr}. Zakharov uses path integrals to get a similar result~~\cite{Zakharov:2000iz}. Similarly, AMY derives an integral equation form of their result~\cite{Jeon:2003gi}. Majumder derives a similar form for photon Bremsstrahlung in the higher twist expansion~\cite{Majumder:2007ne}. All of these approaches are equivalent at their basic level, but they differ in:
\begin{list}{-}{}
 \item The arbitrary form of V(z,q),
 \item The approximations made to derive the explicit form of the operator $\hat{O}(z)$,
 \item The method of solution of the Schrodinger-like equation,
 \item The initial condition (finite time or infinite time) to iterate / evolve,
 \item The implementation of fluctuations in gluon number.
\end{list}
On top of this, there are all the variations in numerical implementation and the treatment of the geometry of nuclear collisions.

\subsection{Solving the Schrodinger equation}
This section has reproduced and repackaged old results, expanding upon (for example) one line of BDMPS which notes a Bethe-Salpeter form of an equation~\cite{Baier:1996kr}. The use of the derivation is in the solution to the Schrodinger equation. The GLV II result gives an all orders in opacity result in closed form. To get the result to arbitrary order, there is no need to go back to the Green's function or to diagrammatic methods, as done for example by Wiedemann~\cite{Wiedemann:2000za}. The result is expressed in a form that can be implemented on computer, in matrices that can be expanded in computer memory as opposed to on paper.

At first, in 2000 when the GLV II paper was written, the all orders result was of limited practical use. First order was quick, second order difficult, third order almost out of reach. Now, with the advances in computing and some numerical optimization of the evaluation, it is feasible to evaluate the result up to and beyond ninth order. Hence, solution to the Schrodinger equation is possible to almost arbitrary precision. Once a form of the equation is settled upon, this method of solution can be compared to other methods, such as an explicit evolution of the quantum wave function with discrete time steps. In this way, the differences between the current set of radiative energy loss formalisms can be resolved.

The numerical solution to the opacity expansion is given in section \ref{sec:rad-allorders}. The qualitative conclusion is similar to that for GLV II: for lengths applicable at RHIC and LHC, the first order is the largest contribution, the higher orders are smaller and cancel each other. This is perhaps surprising: similar to a perturbative expansion, we only expect quick convergence if the driving coupling is small. In this case $L/\lambda$ is not small, it is in the range $1\rightarrow6$. However, there is an additional way in which the second order can be much smaller than the first, one that is not available in introductory quantum mechanics courses. Our potential is imaginary, absorbing. If it is strongly absorbing, then the second order is much smaller than the first, even though $L/\lambda$ is large.

Zakharov noted a similar effect, when radiation was in the `diffusion' zone~\cite{Zakharov:2000iz}. In this region, the BDMPS method of solution, which makes what seems to be a reasonable assumption that if $L/\lambda \gg 1$ then we need to consider $n \gg 1$, in fact misses out the largest contribution, the $n=1$ in opacity. He then goes on to make a different numerical solution to the Schrodinger equation in a subsequent paper~\cite{Zakharov:2004vm}. In future, different numerical solutions to the same equation can be compared.

\subsection{The future: beyond eikonal, small x, static media}
GLV is evaluated in the extreme small $x$ approximation; BDMPS and Zakharov have tried to push past this region. This involves a quantum cascade of both the emitted gluon and the emitting jet in impact parameter space. Relaxing the $x \ll 1$ assumption in the same way for both formalisms will lead to the same Schrodinger equation for both (as long as the color triviality of the result remains).

GLV is evaluated with a Gyulassy-Wang model of static color centers. Moving beyond this involves improved modeling of the $V(q^0,q)$ medium potential, similar to recent work by Djordjevic~\cite{Djordjevic:2007at,Djordjevic:2008iz} or in the AMY formalism (for example, \cite{Jeon:2003gi}). The relaxation of the $q^0 = 0$ assumption may alter the form of the shift operator; however, its remarkably simple form suggests that a simple form may also emerge from a more advanced calculation. In scattering in normal quantum mechanics, the eikonal form emerges out of a more general result: the two-dimensional evolution in impact parameter space is an approximation to the full 3D evolution. The eikonal property is applied to both the jet and the emitted gluon ($k_\perp \ll \omega$). Going beyond this approximation would be valuable.

GLV uses a leading log evaluation of the initial condition, ie the zeroth order in opacity emission. However, there is a significant contribution to the energy loss from the region in which the approximation $\bf{k} \ll x E$ does not apply. Relaxing this assumption will involve relaxing the eikonal, collinear kinematics - remembering that these kinematics have been applied not just to the jet but also to the emitted gluon. AMY evaluates the energy loss for an asymptotic, incoming on-shell jet - to be evaluating the same equation, the evolution from a hard process as in GLV will be needed.

\section{Evaluating the GLV all orders result}
\label{sec:rad-allorders}
\subsection{Summary of formalism}
We turn to the numerical evaluation of the GLV all orders in opacity result. The result in GLV II~\cite{Gyulassy:2000er} is expressed for the case of a massless jet created at a finite time emitting massless gluons. The assumptions and approximations in the derivation include eikonal kinematics, small $x$ energy loss. Djordjevic applied the same formalism but kept explicit mass terms for the jet and emitted gluon, giving the `DGLV' result. This is very similar to the `GLV' result, but with an effective shifted mass term:
\begin{align}
 \beta_{DGLV}^2 &= m_g^2 (1 - x) + M^2 x^2 \\
 \beta_{GLV}^2  &= 0
\end{align}
The final result is expressed in terms of the interference between formation time factors. The inverse of the formation time is the (longitudinal) momentum,
\begin{equation}
 \omega_{(m,\ldots,n)} = \frac{\beta^2 + (\bb{k}-\bb{q}_m - \ldots \bb{q}_n)^2}{2 x E}
\end{equation}
$n$ will always represent the final scatter, $m$ will vary from the the first up to the final scatter. The notation is similar to that in GLV II, with bold letters ($\bb{q},\bb{k}$) representing 2D vectors in the transverse plane. Note that the $\omega_n$ used before is just a special case of this version, with $m=n$. The denominator is the positive light cone momentum of the emitted gluon (assuming $k_\perp \ll \omega$).

The radiated spectrum is expressed in terms of `Cascade' terms:
\begin{equation}
 \bb{C}_{(i_1 i_2 \ldots i_m)} = \frac{(\bb{k}-\bb{q}_{i_1} - \bb{q}_{i_2} - \ldots - \bb{q}_{i_m} )}{ \beta^2 + (\bb{k}-\bb{q}_{i_1} - \bb{q}_{i_2} - \ldots - \bb{q}_{i_m} )^2 }
\end{equation}
with $\beta$ being either $\beta_{GLV}$ or $\beta_{DGLV}$ from above. This represents the shifting of the momentum of the radiated gluon due to momentum kicks from the medium. We define the `Hard' term as a special case of $\bb{C}$ without any momentum shifts:
\begin{equation}
 \bb{H} = \frac{\bb{k}}{\beta^2 + \bb{k}^2}
\end{equation}
The `Gunion-Bertsch' scattering terms are linear combinations of $\bb{C}$:
\begin{equation}
 \bb{B}_{(i_1 i_2 \ldots i_m)(j_1 j_2 \ldots j_m)} = \bb{C}_{(i_1 i_2 \ldots i_m)} - \bb{C}_{(j_1 j_2 \ldots j_m)}
\end{equation}
with the special case
\begin{equation}
 \bb{B}_i = \bb{H} - \bb{C}_i
\end{equation}

The full result, differential with respect to $x$ and $\bb{k}$ (which is uniform over $\phi_k$), for the radiation spectrum at nth order in opacity is expressed as
\begin{align}
\label{eqn:allorders}
 x&\frac{dN^{(n)}}{dx\, d^2 {\bf k}} =
\frac{C_R \alpha_s}{\pi^2} \frac{1}{n!} 
\int \prod_{i=1}^n \left( 
\frac{L d^2{\bf q}_{i}}{\lambda_g(i)} 
\left[\bar{v}_i^2({\bf q}_{i}) - \delta^2({\bf q}_{i}) \right]\right) \nonumber \\
& \times \Bigl( -2\,{\bf C}_{(1, \cdots ,n)} \cdot 
\sum_{m=1}^n {\bf B}_{(m+1, \cdots ,n)(m, \cdots, n)} \times  \nonumber \\
& \left[ \cos \sum_{k=2}^m \omega_{(k,\cdots,n)} \Delta z_k
-   \cos \sum_{k=1}^m \omega_{(k,\cdots,n)} \Delta z_k 
\right] \Bigr)
\end{align}
where $\sum_2^1 \equiv 0$ is understood and $|\bar{v}_i({\bf q}_{i}) |^2$ is 
defined as the normalized distribution of momentum transfers from 
the $i^{{\rm th}}$ scattering center, $\lambda_g(i)$ is the mean free path of the emitted gluon, $C_R$ is the color Casimir of the jet. 

Noting the discussion in section \ref{sec:formulaglvii} about classical vs quantum cascades, the expression for the source term (the only term that changes the energy distribution) at all orders in opacity is:
\begin{align}
\label{eqn:allorders2}
 x & \frac{dN^{(n)}}{dx\, d^2 {\bf k}} =
\frac{C_R \alpha_s}{\pi^2} \frac{1}{n!} 
\int \prod_{i=1}^n \left(
\frac{L d^2{\bf q}_{i} }{\lambda_g(i)} 
\left[\bar{v}_i^2({\bf q}_{i}) - \delta^2({\bf q}_{i}) \right]\right) \nonumber \\
& \times \Bigl( -2\,{\bf C}_{(1, \cdots ,n)} \cdot {\bf B}_{n} \times \nonumber \\
& \left[ \cos \sum_{k=2}^n \omega_{(k,\cdots,n)} \Delta z_k
-   \cos \sum_{k=1}^n \omega_{(k,\cdots,n)} \Delta z_k
\right]\; \Bigr)
\end{align}

\subsubsection{Smooth background density}
For evaluation in a smooth background density, we change
\begin{align}
\frac{1}{n!} &
\int \prod_{i=1}^n \left( d^2{\bf q}_{i}  \, \frac{L}{\lambda_g(i)} \left[\bar{v}_i^2({\bf q}_{i}) - \delta^2({\bf q}_{i}) \right] \right) \rightarrow \nonumber \\
 \frac{1}{n!} &  \int_0^\infty dz_1 \cdots \int_{z_{n-1}}^\infty dz_n \int \prod_{i=1}^n \left( d^2{\bf q}_{i}\, \frac{ \bar{v}^2({\bf q}_{i})- \delta^2({\bf q}_{i}) }{\lambda(z)} \right)
\end{align}
where $\lambda(z)$ is is the distance (and time) dependent mean free path. 

\subsubsection{Static color center model}
The GLV II result is not confined to a specific model of the jet-medium interaction. The $\delta(q^0)$ in the Gyulassy-Wang potential has been used in the derivation, but work by Djordjevic and Heinz suggests that this strong assumption may not be needed to give the same form of result~\cite{Djordjevic:2008iz} (with a slightly modified potential). In an HTL calculation, the unscreened magnetic interaction sends the mean free path to zero, violating the assumption $\lambda \gg 1/\mu$. However, these soft interactions do not contribute to much of anything, except collision number. An ad-hoc magnetic screening $\mu_{mag} \approx \mu_D$ pushes the mean free path up to $\approx 1fm$ again, without strongly affecting the $\av{\Delta p}$ or $\av{q_\perp^2}$.

For this evaluation, we stay with the simpler Gyulassy-Wang static color center model. For a specific $\omega, k$, the mean free path does not affect the formation time. Hence, if the opacity series peaks at $n=1$ for the static model (from Zakharov, if $\tau_f \gg L$), this should persist to other models with shorter mean free paths.

For the Yukawa screened interactions, the differential gluon cross section in the local density approximation is 
\begin{align}
\sigma(z_i,{\bf q}_i) &\equiv \sigma_{el}(z_i)|\bar{v}_i({\bf q}_{i})|^2=
\frac{d^2 \sigma_{el}(z_i)}{d^2{\bf q}_{i}} \nonumber \\
 & =\frac{\sigma_{el}(z_i)}{\pi}
\frac{\mu(z_i)^2}{({\bf q}^{2}+\mu(z_i)^2)^2}
\end{align}
where $\mu(z_i)$ is the local Debye mass (that may vary if the system expands). 

The total cross-section, $\sigma_{el}$ is related to the (gluon) mean free path, $\lambda = 1/(\rho \sigma_{el})$ where $\rho$ is the local density, and is given by
\begin{equation}
 \sigma_{el} = \frac{\pi  C_R C_A \alpha^2}{2\mu^2}
\end{equation}

\subsubsection{Relating $\lambda$ to $\hat{q}$}
\label{sec:madasaqhatter}
The average $\bb{q}$ kick per collision, for $q_{max} \approx \sqrt{6 T E} \gg \mu$, given by this potential is $\av{\bb{q}} = \bb{0}$, but with
\begin{equation}
 \av{\bb{q}^2}_{\mbox{per collision}} = \mu^2 \log \frac{E^2}{\mu^2}
\end{equation}
hence the average $\bb{q}$ per unit distance is
\begin{equation}
 \av{\bb{q}^2}_{\mbox{per fm}} = \frac{\mu^2}{\lambda} \log \frac{E^2}{\mu^2} = \hat{q} \log \frac{E^2}{\mu^2}
\end{equation}
We could present the result in terms of $\hat{q}$ (which here $\ne \av{q_\perp^2}$) by the substitution
\begin{equation}
 \frac{1}{\lambda} = \frac{\hat{q}}{\mu^2}
\end{equation}
and expressing in terms of this $\hat{q}$, but we still have a dependence on $\mu$. In general, $\hat{q}$ does not uniquely give the energy loss; you either have to specify both $\hat{q}$ and $\mu$, or both $\lambda$ and $\mu$. In a kinetic theory, $\lambda$ and $\mu$ are related to the temperature. Here, we will set $\lambda = 1$fm, $\mu = 0.5$ GeV.

\subsection{The uncorrelated medium}
In general, $\lambda(z)$ can vary with $z$. But even if we have a box of constant temperature, $\lambda(z) = \lambda$, the spacing of collisions is mutually dependent. Taking the example of second order in opacity calculation confined to a length L, if the first collision happens at 0.9L, the spacing to the second is limited to less than 0.1L.

A significant numerical simplification is achieved by removing this mutual dependence, and evaluating an `uncorrelated medium'. This can be expressed as~\cite{Gyulassy:2000er}
\begin{equation}
\label{eqn:uncorr}
\bar{\rho}(z_1,\cdots,z_n)=  \prod_{j=1}^n
\frac{\theta(\Delta z_j)}{L_e(n)}e^{-\Delta z_j/L_e(n)} \;\;,
\end{equation}
The spacings of collisions are independent. This can only be considered an approximation to a box profile, which is then only an approximation to the varying density profile. The full numerical result is not limited to this approximation, and future work should examine the difference between the box approximation and the uncorrelated medium approximation.

\eq{eqn:uncorr} converts the oscillating formation physics factors in the full result into simple Lorentzian factors
\begin{equation}
\int d\bar{\rho}
\cos \sum_{k=j}^m \omega_{(k,\cdots,n)} \Delta z_k
={\rm Re}\;\prod_{k=j}^m \frac{1}{1+i\omega_{(k,\cdots,n)}L_e(n)}
\; \; , 
\end{equation} 
In order to fix $L_e(n)$, we require that $\langle z_k-z_0 \rangle =k L/(n+1)$ for both geometries. This constrains $L_e(n)=L/(n+1)$.

\subsection{Monte Carlo evaluation}
The GLV result is ideally suited to Monte-Carlo evaluation. For high orders in opacity, it is a high dimensional integral (either 2n or 3n depending on whether the uncorrelated medium approximation is used or not) and should not be dramatically peaked in one small region (although it is highly oscillatory). 

Monte Carlo integration proceeds by repeated evaluation of the integrand. By throwing dice as to the value of the variables (over which the integration is being evaluated) and averaging the results, an approximation is made to the result of the integral. For large numbers of evaluations, the result should converge to the true value of the integral as $1/\sqrt{n}$. The result is expressed as~\cite{NumRecipesC}
\begin{equation}
 \int f dV \approx V \av{f} \pm V \sqrt{\frac{\av{f^2}-\av{f}^2}{N}}
\end{equation}
where
\begin{equation}
\label{eqn:monte}
 \av{f} = \frac{1}{N}\sum^N_{i=1} f(x_i) \mbox{   } \av{f^2} = \frac{1}{N} \sum^N_{i=1} f^2(x_i)
\end{equation}
In this way, an estimate of the result and a rough estimate of the uncertainty on the result is found. The use of quasi-random numbers can often improve the rate of convergence of Monte-Carlo integration, but this is not pursued further here.

The key to the multi-order implementation is the numerical evaluation of vector dot products between $C$ and $B$. Considerable effort is saved by iterating these numerically as opposed to by hand on paper. The results presented here use only the most basic non-adaptive Monte-Carlo integrators. This study is presented as a starting point, to show that high orders in opacity are well within the reach of current computing power. 

Only simple fixed length results will be shown. The gluon mean free path is fixed at $\lambda_g = 1$fm, the Debye mass at $\mu_D = 0.5$GeV. The $q_\perp$ momentum integral is cut at $\sqrt{6 T E}$, with T=0.25GeV.

\section{All orders results}
\label{sec:allordersresults}
Results were presented in GLV II \cite{Gyulassy:2000er} for summation up to 1st, 2nd and 3rd orders in opacity. In appendix \ref{sec:comp}, explicit comparisons to some of the GLV II results are shown. Good consistency is found for first and second orders, while the third order result for \cite{Gyulassy:2000er} is a little smaller than that found here. Error bars on all results are shown, as per the uncertainty estimate from \eq{eqn:monte}. 

Results plotted in this chapter are for the one-gluon distributions. In order to produce an $\raa(p_T)$ result, one gluon distributions are usually input into a Poisson convolution before being combined with initial spectra. Many of the results are compared to the Bertsch-Gunion spectrum, the incoherent radiation spectrum with a Gyulassy-Wang model and a small 0.1 GeV screening of the $\bb{k_\perp} = \bb{q_\perp}$ pole.

\subsection{Results - $dN/dxdk$}
\subsubsection{p = 100 GeV}
The interpretation of the results is complicated by the effect of kinematic limits, so we start with results for light quarks with $p_T$=100GeV. The individual orders in opacity (up to n=9) are presented in appendix section \ref{sec:e100l6}, for L=6fm, E=100GeV, for $x=0.025, 0.05, 0.2, 0.5, 0.8$, for both the full result (including the classical cascading) and for only the quantum source term (the part that is important for dN/dx). Here, we concentrate on the answers summed up to a specific order. Unless otherwise stated, a result shown uses the full result including the classical cascade, ie \eq{eqn:allorders}.

To summarize: the first order result gives a good approximation to the all orders result for large $k_\perp$, short lengths, large $x$. 

For all the dN/dxdk results (shown in \fig{plot:radNth1}, \fig{plot:radNth2}, \fig{plot:radNth3}, \fig{plot:radNth4}), the all orders (summed up to 9th order) result deviates from the first order for small $k_\perp$, and then follows the first order result for large $k_\perp$. The point of reattachment depends on $x$, the energy of the emitted gluon. This can be seen in \fig{plot:radNth4}: for x=0.05 (for a 100 GeV jet), this point is $k_\perp \approx 3.5$GeV, for x=0.5 the point moves up to $k_\perp \approx 7$GeV.

Below this point, the behavior depends on the length, as seen in \fig{plot:radNth3}, and on the emitted gluon energy (ie $x$) as seen in \fig{plot:radNth2}. For short lengths, L=1fm (with the gluon mean free path $\lambda$=1fm as well), the higher orders make little difference. For longer lengths, the higher orders suppress the first order peak, and also suppress the small negative dip for very small $k_\perp \approx 0.5$GeV. 

This is the effect of interference between radiation from multiple scattering. The first order result only has the interference with the creation radiation. As the length increases, the first order result pushes its collision out to an increasing $L/2$, the interference with the creation point becomes less severe and the result closer to incoherent. The interference between multiple collisions comes into play and keeps the result further from the incoherent.

For a specific $k_\perp$, the result increases with increasing L, even when weighed by $1/L$ as in \fig{plot:radNth3}. Hence, the length dependence of the average must be stronger than $\propto L$. However, the higher orders have more of an effect on the result for longer lengths. The full result grows (slightly) more slowly than the first order result. The full result should approach a linear growth (ie a common curve when weighed by $1/L$) for very long lengths.

The opacity series converges slowly for small $k_\perp$ and small x. In the appendix section \ref{sec:e100l6}, the opacity series is seen to be oscillatory, with large successive terms almost canceling each other - for small x. For large x, similar shapes in $k_\perp$ are seen, just with the higher orders dramatically decreasing in magnitude. In \fig{plot:radNth1}, even the fifth order result does not give a good result compared to the ninth order result for the shape of the small $k_\perp$ spectrum, for x=0.05. This can also be seen in \fig{plot:radNth4}, where the quantum source term is plotted, removing the classical cascade. For x=0.05, the first order is a good approximation down to $k_\perp\approx 4$GeV, summing up to fifth order is a good approximation down to $k_\perp \approx 1$GeV. For x=0.5, the fifth order is a good approximation all the way down to $k_\perp \approx 0$.

For large $k_\perp$, lengths $L=1 \rightarrow 4$fm, and (for a 100GeV jet) $x > 0.05$, the first order result is dominant. As $k_\perp$ is pushed down, suppression from the first order result is seen. We will see the effect of this is the next section, when plotting $dN/dx$, the input into an $\raa(p_T)$ calculation.

\subsubsection{p = 20GeV}
The use of a 100GeV jet has been convenient here, to remove (ie push up in $k_\perp$) the cuts of kinematic limits, $k_{max} = 2 x (1-x) E$. x=0.05 for a 100GeV jet is a gluon of 5GeV. The problematic nature of radiative energy loss calculations for lower energy jets is illustrated in \fig{plot:rade20}: x=0.05 for a 20 GeV jet corresponds to a 1 GeV emitted gluon, far from hard relative to the medium (remember, both the jet and the emitted gluon are considered to be eikonal). The $k_{max}$ cut in this region is so harsh, that we might consider the remainder of the calculation moot. While this may not be important for the calculation of the average energy loss, it is important for $\raa(p_T)$, which is more sensitive to the distribution in number of gluons, which is peaked at small $x$.

A cut on a distribution is a usually a sign of pushing kinematic approximations too far away from their region of applicability. The hope is that little of the distribution sits in the region of $k_{max}$, so it could be varied up and down without changing the result. For jets at RHIC, this just is not possible.

\subsection{Results - DGLV vs GLV}

The results in the previous section were given for the GLV result, with the mass of the jet and the emitted gluon as zero. For consistency, we would hope that the DGLV result would give something similar for small jet and gluon masses. But the small mass of the emitted gluon has a surprisingly large effect, as shown for first order in \fig{plot:rade20dglvn1} and summed up to ninth order in \fig{plot:rade20dglvn9}. The result is changed well beyond an ad-hoc cut on the GLV spectrum at $k_\perp = \mu$. This is worrying for the accuracy of the results. It might be necessary to include better considerations of the in-medium dispersion relation of the emitted gluon.

\begin{figure*}[p!] 
\centering
$\begin{array}{c}
\epsfig{file=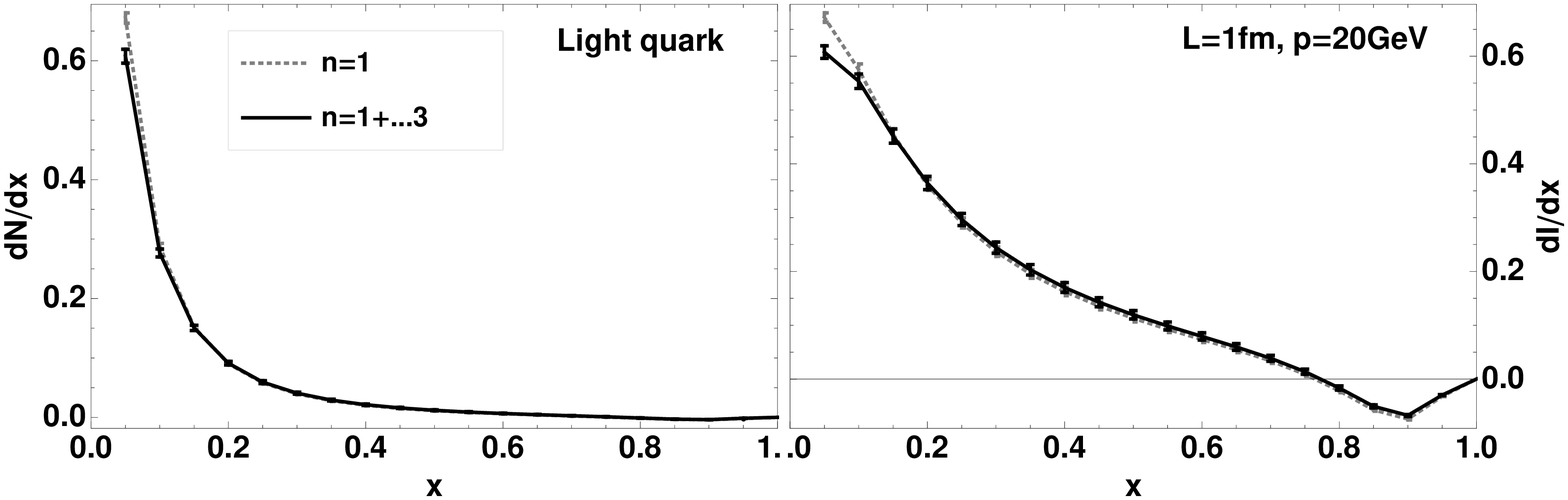,width=5in,clip=,angle=0} \\
\epsfig{file=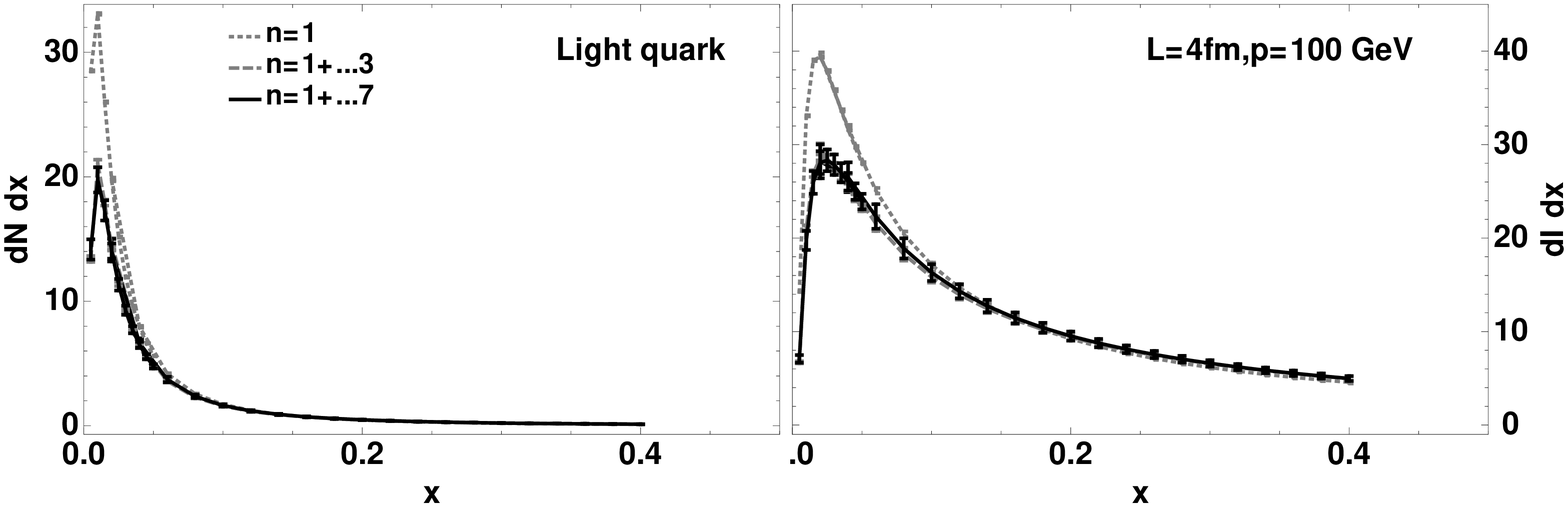,width=5in,clip=,angle=0}
\end{array}$
\caption{\label{plot:raddNdxL1} 
The one-gluon number (dN/dx - left) and energy distribution (dI/dx -right) of radiation using DGLV. The upper plot shows the short length L=1fm for a 20 GeV jet; the lower plot shows a longer length L=4fm for a 100 GeV jet. For both, higher orders than those shown are negligible.
}
\end{figure*}

\begin{figure*}[p!] 
\centering
$\begin{array}{c}
\epsfig{file=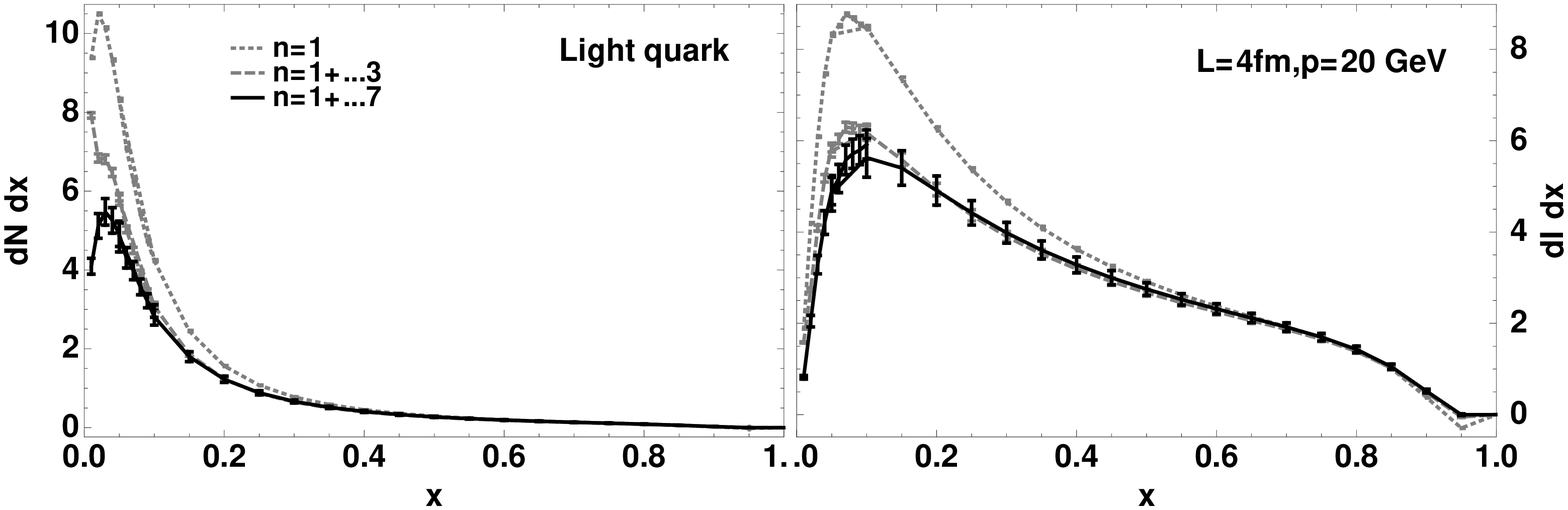,width=5in,clip=,angle=0} \\
\epsfig{file=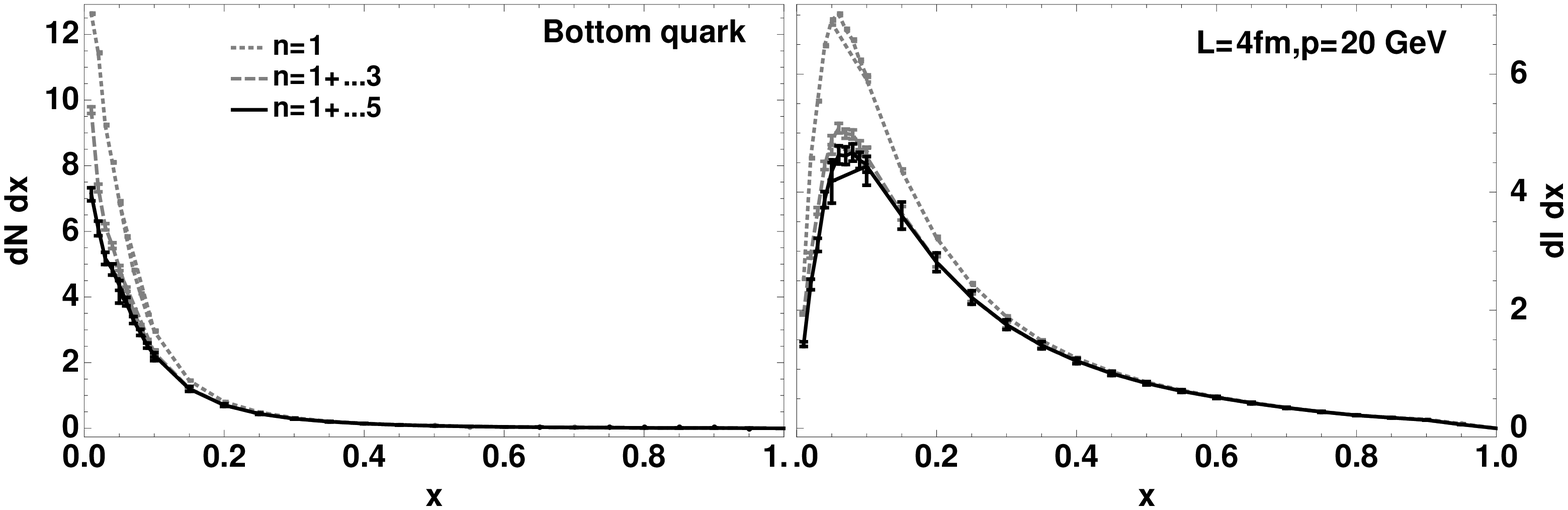,width=5in,clip=,angle=0}
\end{array}$
\caption{\label{plot:raddNdxL3} 
The one-gluon number (dN/dx) and energy distribution (dI/dx) for a light quark and a bottom quark jet with momentum 20 GeV, with L=4fm, at first order and summed up to higher orders.
}
\end{figure*}

\subsection{Results - $dN/dx$ and $dI/dx$}

For calculations of $R_{AA}(p_T)$, the important quantity is jet energy loss and the important distribution is $dN/dx$. For the calculation of the average energy loss, the relevant distribution is $dI/dx = xE dN/dx$. As the Gyulassy-Wang potential has an explicit $\delta(q^0)$ factor, the energy of the emitted gluon equals the energy loss of the jet (if this factor is removed, a more considered translation between energy emitted and energy loss is needed, due to possible energy transfers in the collision).

In order to avoid the details of a $k_T > \mu$ cut, the results in this section use the DGLV formula, including a small mass for the jet and emitted gluon - despite the reservations expressed in the previous section. As before, the jet propagates through a medium of fixed density, fixed length, with a gluon mean free path $\lambda = 1 fm$ and a Debye mass $\mu = 0.5 GeV$. The cut on large $k_\perp$ is also important - here the value given by Djordjevic~\cite{Djordjevic:2003zk} is used, $k_{\perp,max} = 2 x (1 - x ) E$.

The figures (\fig{plot:raddNdxL1}, \fig{plot:raddNdxL3}) show that the first order in opacity dominates the results. For lengths up to L=4fm, the second and third order provide small alterations on top of this, reducing the number of gluons emitted at small x. The orders above third order are negligible in considerations of $dN/dx$.

\section{Conclusions}

The opacity expansion is a Dyson expansion of a Schrodinger-like equation. It provides both a derivation of that Schrodinger equation and a method of solution to almost arbitrary accuracy. GLV II gave a closed form expression for the all orders result. This result can be decomposed into a quantum source term and a classical cascade of emitted gluons.

A new numerical evaluation of the GLV all orders result has been presented. Monte-Carlo integration is well suited to the high order integrals, and a basic version has been implemented to give results up to ninth order in opacity. These results give qualitatively the same conclusions as in GLV II: for the lengths and energies of interest, the first order in opacity is the largest contribution. The higher orders reduce the gluon emission for longer lengths and small $k_\perp$. This small $k_\perp$ region change affects gluons below $\approx 4$ GeV. This has little effect on the average energy loss of jets above $\approx 20$ GeV. $\raa(p_T)$ is dependent on the number distribution after a multi-gluon emission convolution, so will be affected more than the average energy loss.

Further investigation is needed to give full $\raa(p_T)$ results (and other observables). It is possible that the inclusion of higher orders may have an effect differential in mass, as coherence is affected by the mass of the jet. Work is also needed on the evaluation without the uncorrelated medium assumption, and on the result in a medium of arbitrary varying density. This calculation is numerically more intensive, but is still possible with today's computing power.

The difference between the GLV and DGLV distributions in dN/dxdk for a light quark jet is worrying. The inclusion of a small gluon mass, $m_g \approx 0.3$GeV, makes a surpisingly large difference.  Further analysis of the medium effects on gluons with $\omega \lesssim 3$GeV, $k_\perp \lesssim 3$GeV is needed. 

With numerical optimization and greater computing time, orders in opacity beyond ninth are easily accessible. While this is likely not important for RHIC or LHC applications (for perturbative input parameters such as $\lambda \approx 1$fm), it should provide important insight into the phenomenon of coherent radiation in thick media, and possible contact with the approximations made in the evaluation of the BDMPS result.

Further improvements to the formalism are necessary. The derivation in GLV II applies to small x, eikonal kinematics and no energy exchange with the medium. Pushing beyond these assumptions is likely necessary for quantitative work, especially for emitted gluons of 1-2 GeV. The eikonal assumption applies both to the jet and the emitted gluon, a poor assumption in the region of interest. Despite its complicated derivation from QCD diagrams, the GLV II operator recursion from order to order in opacity is remarkably simple, and corresponds to a remarkably simple Schrodinger-like equation (as in BDMPS and Zakharov's work). There is a tempting substitution of the 2D, transverse vectors into 3 or 4D vectors, the longitudinal momentum shift into a general energy shift for an arbitrary potential. However, much work is needed to prove or disprove a simple result for the more general kinematics that are desired.

{\em Acknowledgments: Valuable input from Miklos Gyulassy is gratefully acknowledged, as well as discussions with Ivan Vitev.}

\onecolumngrid
\clearpage

\begin{appendix}
 
\section{Comparison to GLV II}
\label{sec:comp}

\begin{figure*}[htb] 
\centering
$\begin{array}{ccc}
\epsfig{file=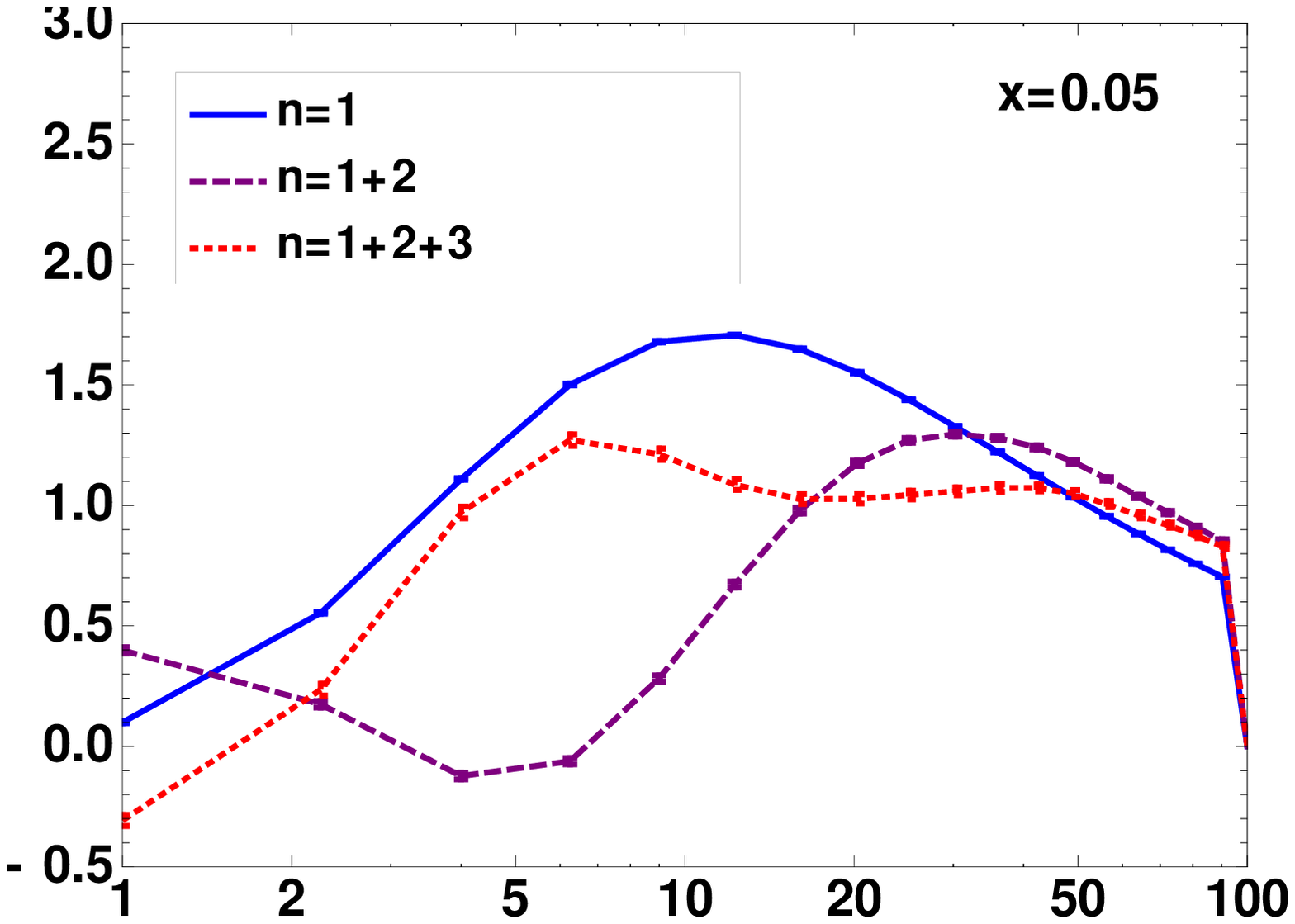,width=2.in,clip=,angle=0} &
\epsfig{file=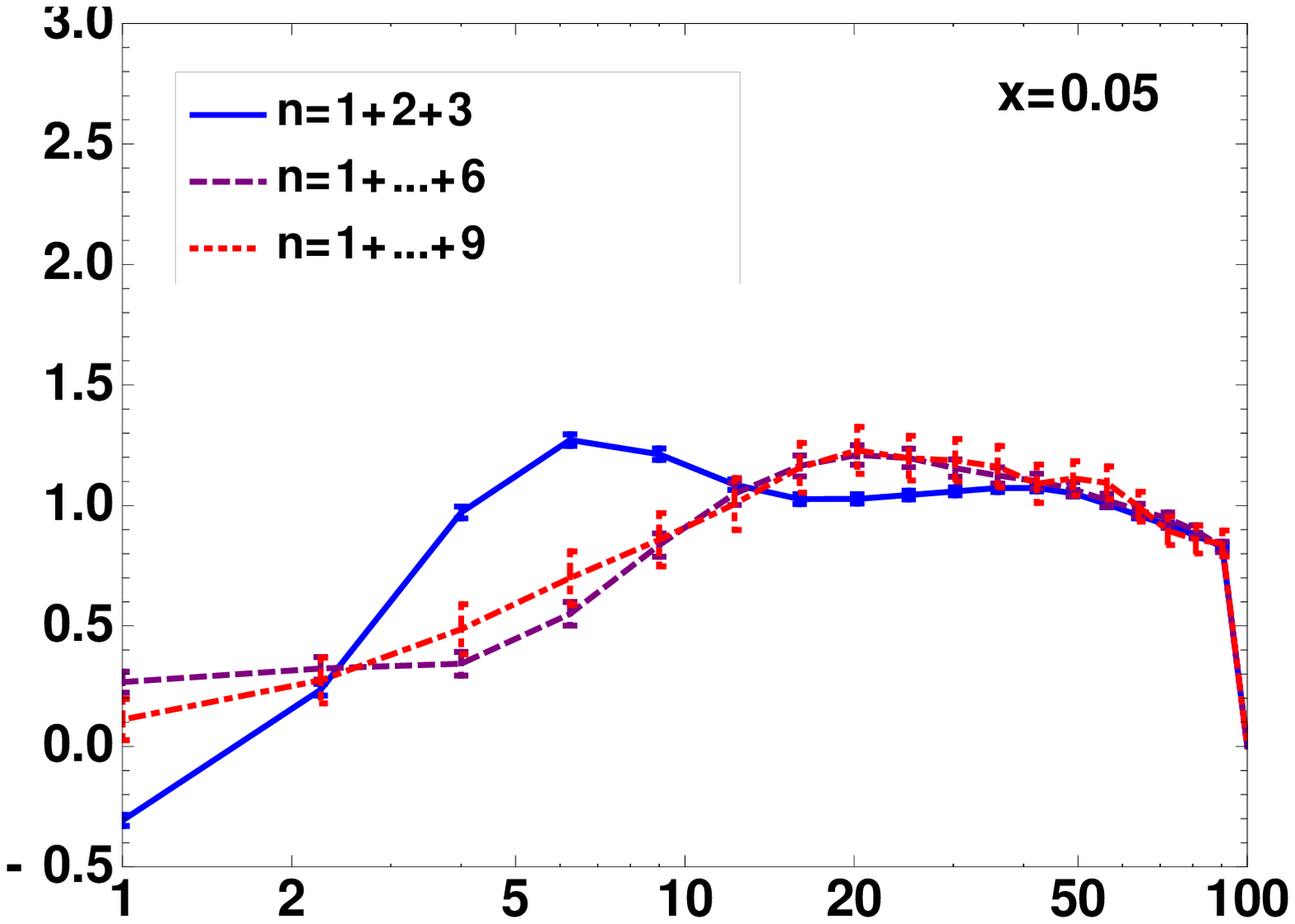,width=2.in,clip=,angle=0} &
\epsfig{file=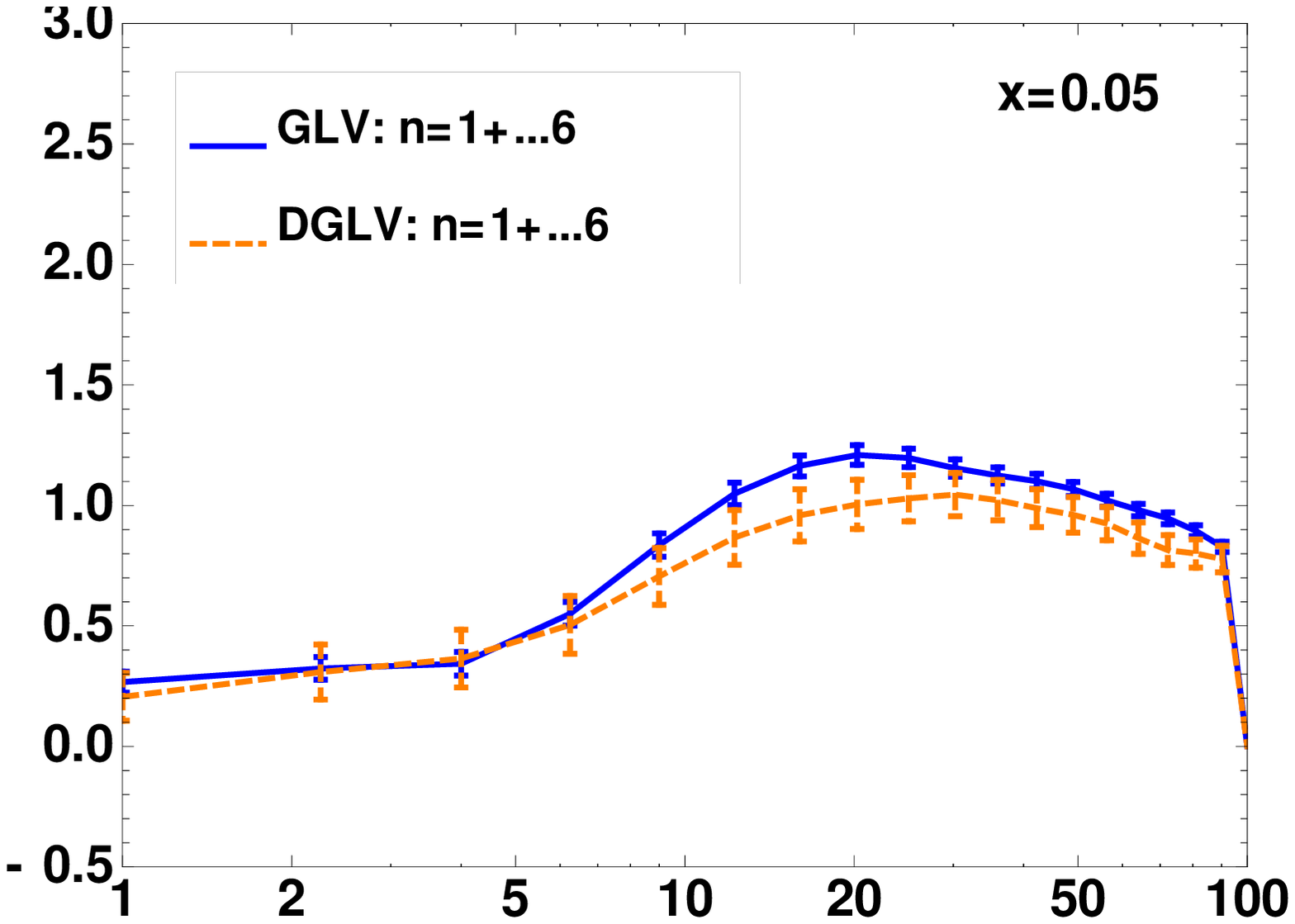,width=2.in,clip=,angle=0}
\end{array}$
\caption{\label{plot:ivancheck1} 
Results from the new Monte-Carlo evaluation, to be compared to fig 4 from GLV II~\cite{Gyulassy:2000er}, for L=5fm, x=0.05 of a 100GeV jet. The left plot shows the result for first, first plus second, and summed up to third orders; the middle and right show the effect of orders up to ninth order (left) and the DGLV inclusion of a jet and gluon mass (right).
}
\end{figure*}

\begin{figure*}[htb] 
\centering
$\begin{array}{ccc}
\epsfig{file=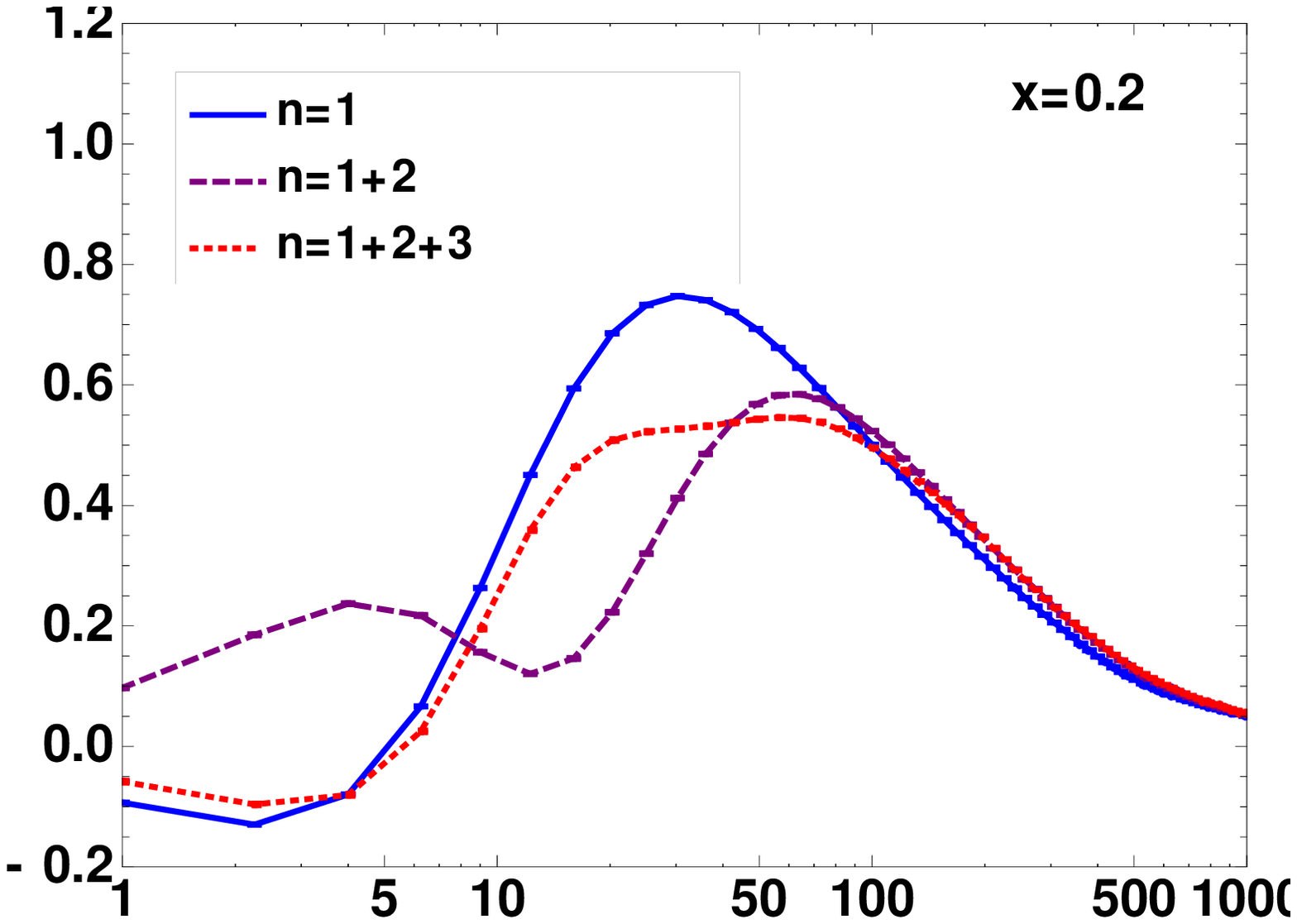,width=2.in,clip=,angle=0} &
\epsfig{file=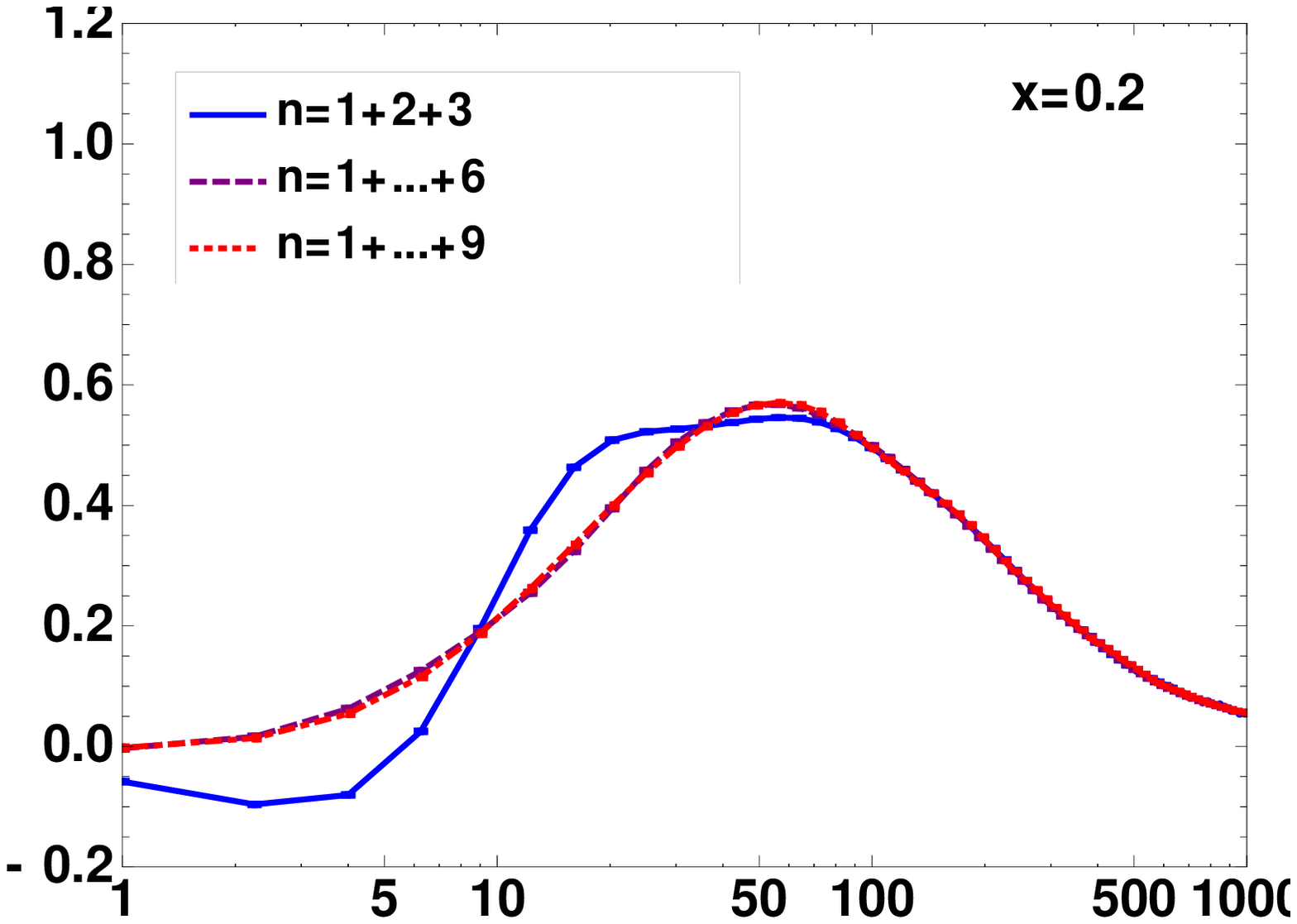,width=2.in,clip=,angle=0} &
\epsfig{file=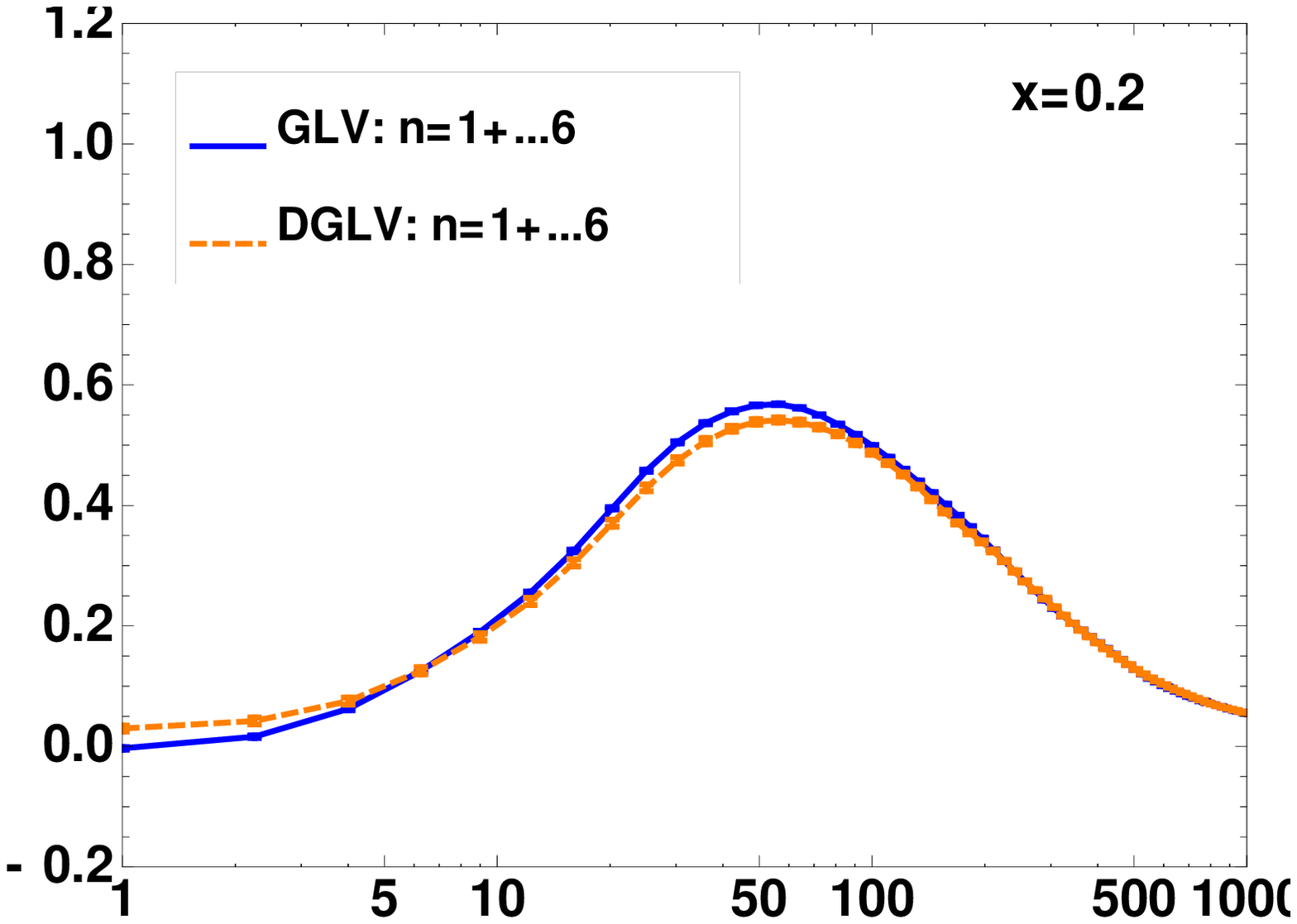,width=2.in,clip=,angle=0}
\end{array}$
\caption{\label{plot:ivancheck2} 
Results from the new Monte-Carlo evaluation, to be compared to fig 4 from GLV II~\cite{Gyulassy:2000er}, for L=5fm, x=0.2 of a 100GeV jet. The left plot shows the result for first, first plus second, and summed up to third orders; the middle and right show the effect of orders up to ninth order (left) and the DGLV inclusion of a jet and gluon mass (right).
}
\end{figure*}

\begin{figure*}[htb] 
\centering
$\begin{array}{ccc}
\epsfig{file=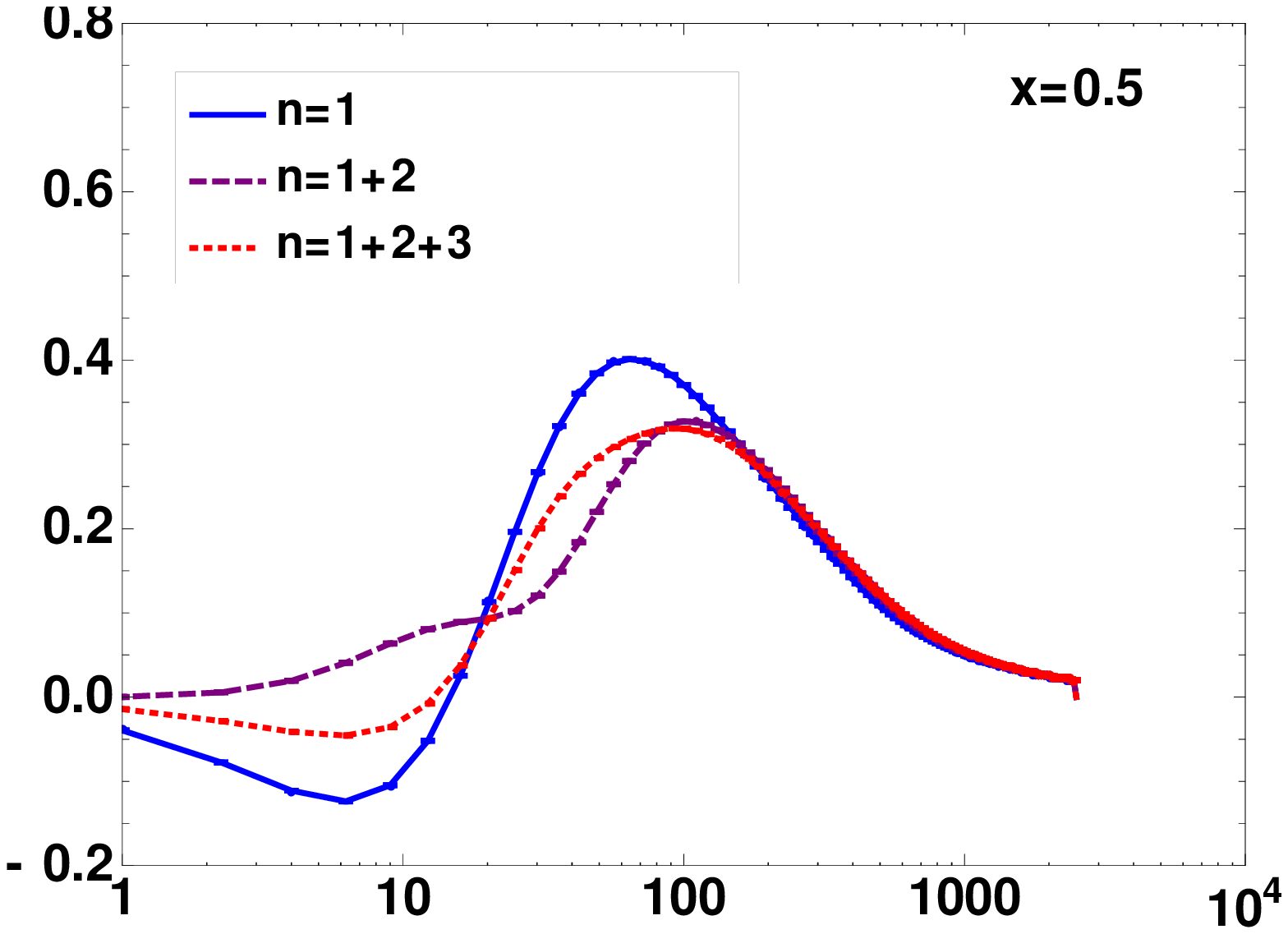,width=2.in,clip=,angle=0} &
\epsfig{file=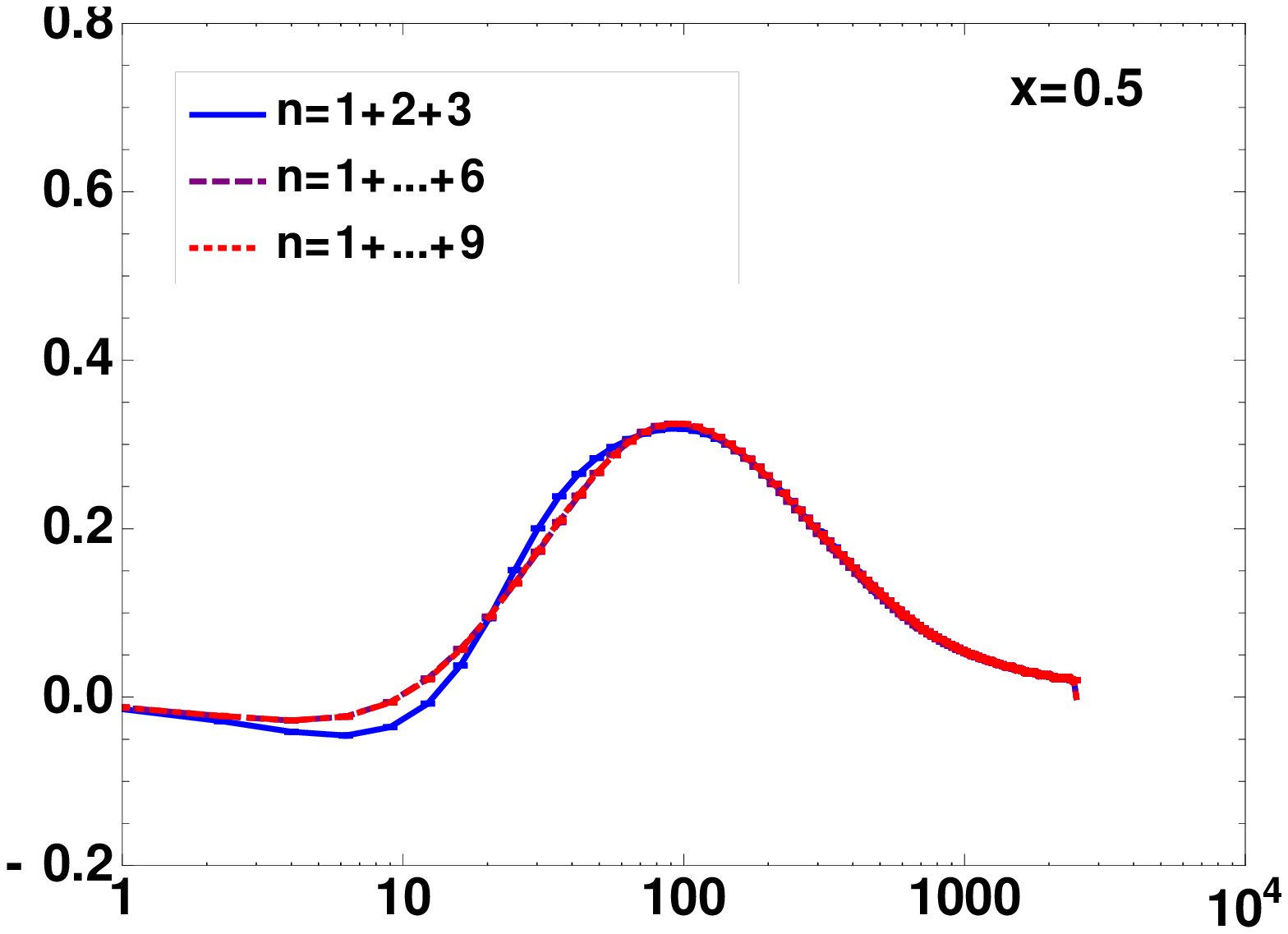,width=2.in,clip=,angle=0} &
\epsfig{file=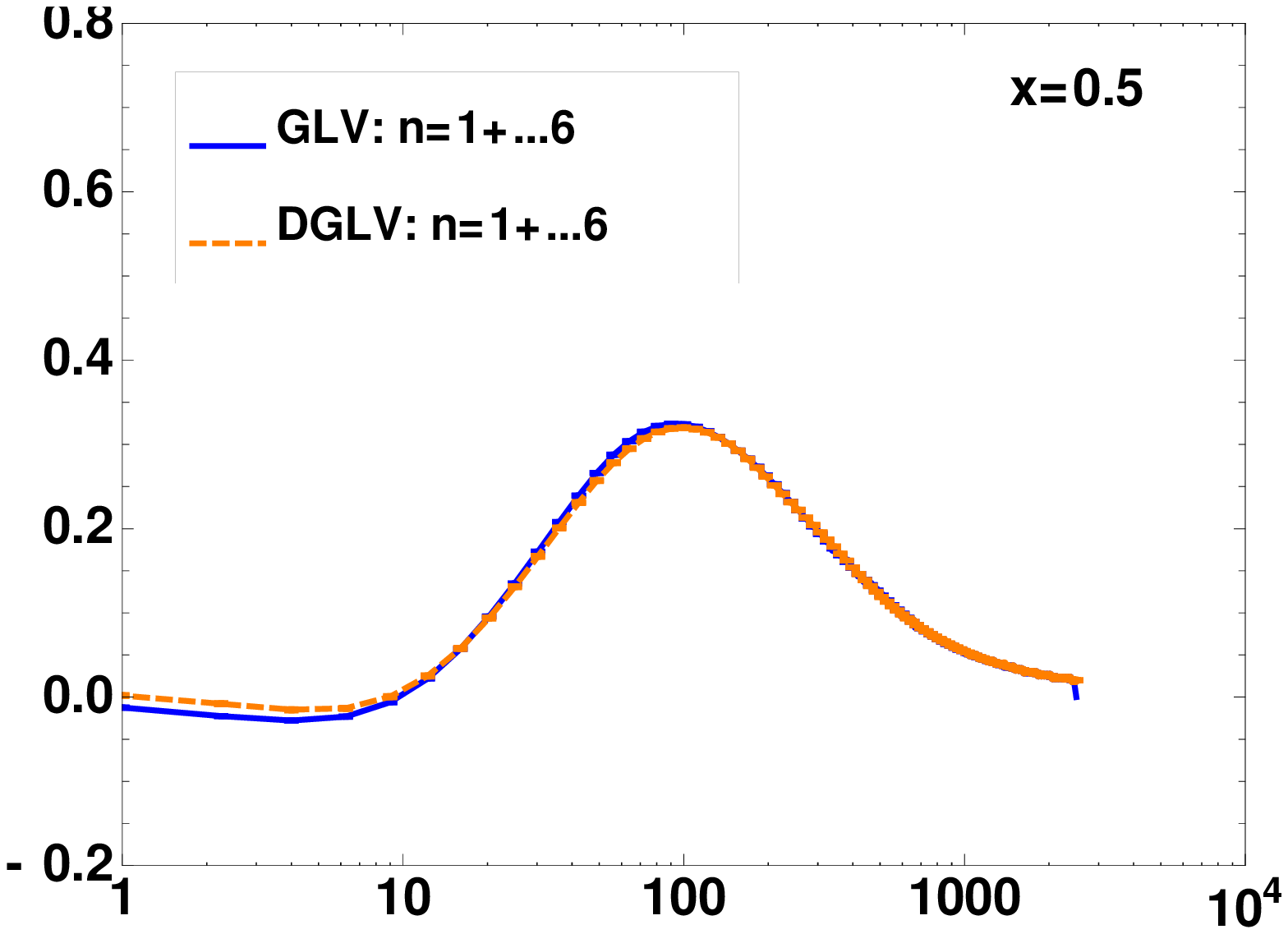,width=2.in,clip=,angle=0}
\end{array}$
\caption{\label{plot:ivancheck3} 
Results from the new Monte-Carlo evaluation, to be compared to fig 4 from GLV II~\cite{Gyulassy:2000er}, for L=5fm, x=0.5 of a 100GeV jet. The left plot shows the result for first, first plus second, and summed up to third orders; the middle and right show the effect of orders up to ninth order (left) and the DGLV inclusion of a jet and gluon mass (right).
}
\end{figure*}

\clearpage
\section{Results for L=6fm, n=1...9}

\label{sec:e100l6}
\begin{figure*}[htb] 
\centering
$\begin{array}{cc}
&
\epsfig{file=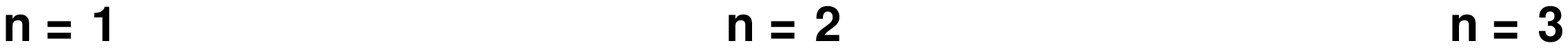,width=4.8in,clip=,angle=0} \\
\epsfig{file=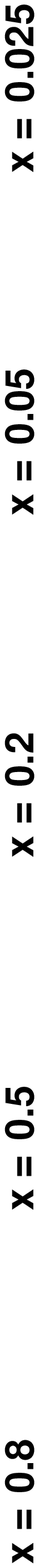,width=0.2in,clip=,angle=0} &
\epsfig{file=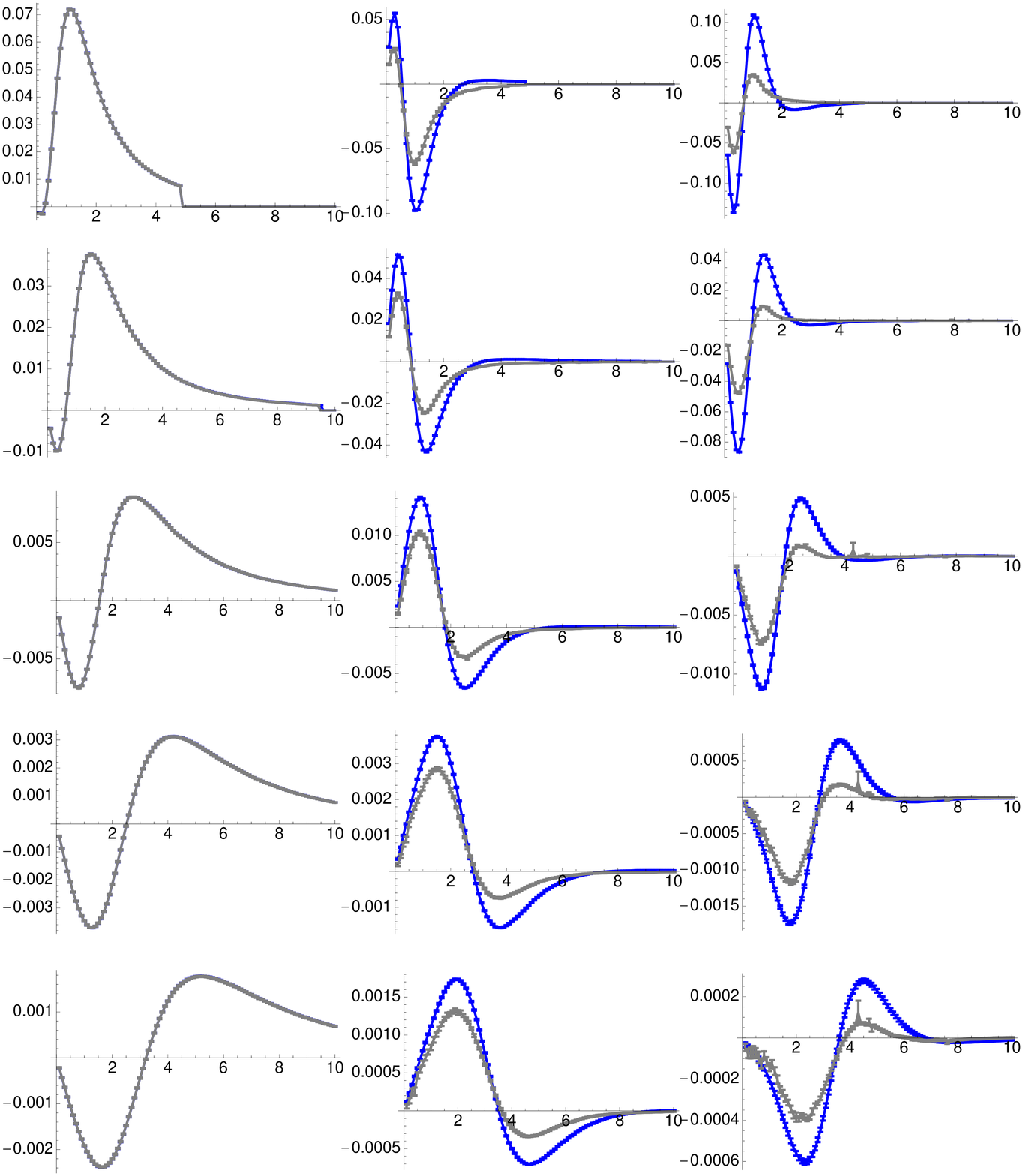,width=4.8in,clip=,angle=0}
\end{array}$
\caption{\label{plot:e100n123} 
The dN/dxdk GLV results for individual orders in opacity 1,2,3 for a 100 GeV massless quark jet. Shown is the result for the full result including the classical cascade, and the result only for the quantum source term. 
}
\end{figure*}

\begin{figure*}[htb] 
\centering
$\begin{array}{cc}
&
\epsfig{file=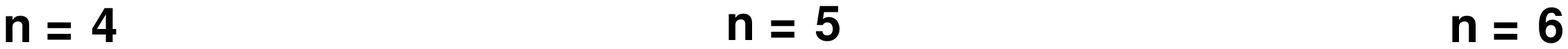,width=4.8in,clip=,angle=0} \\
\epsfig{file=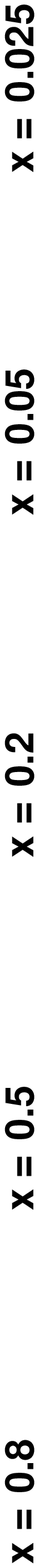,width=0.2in,clip=,angle=0} &
\epsfig{file=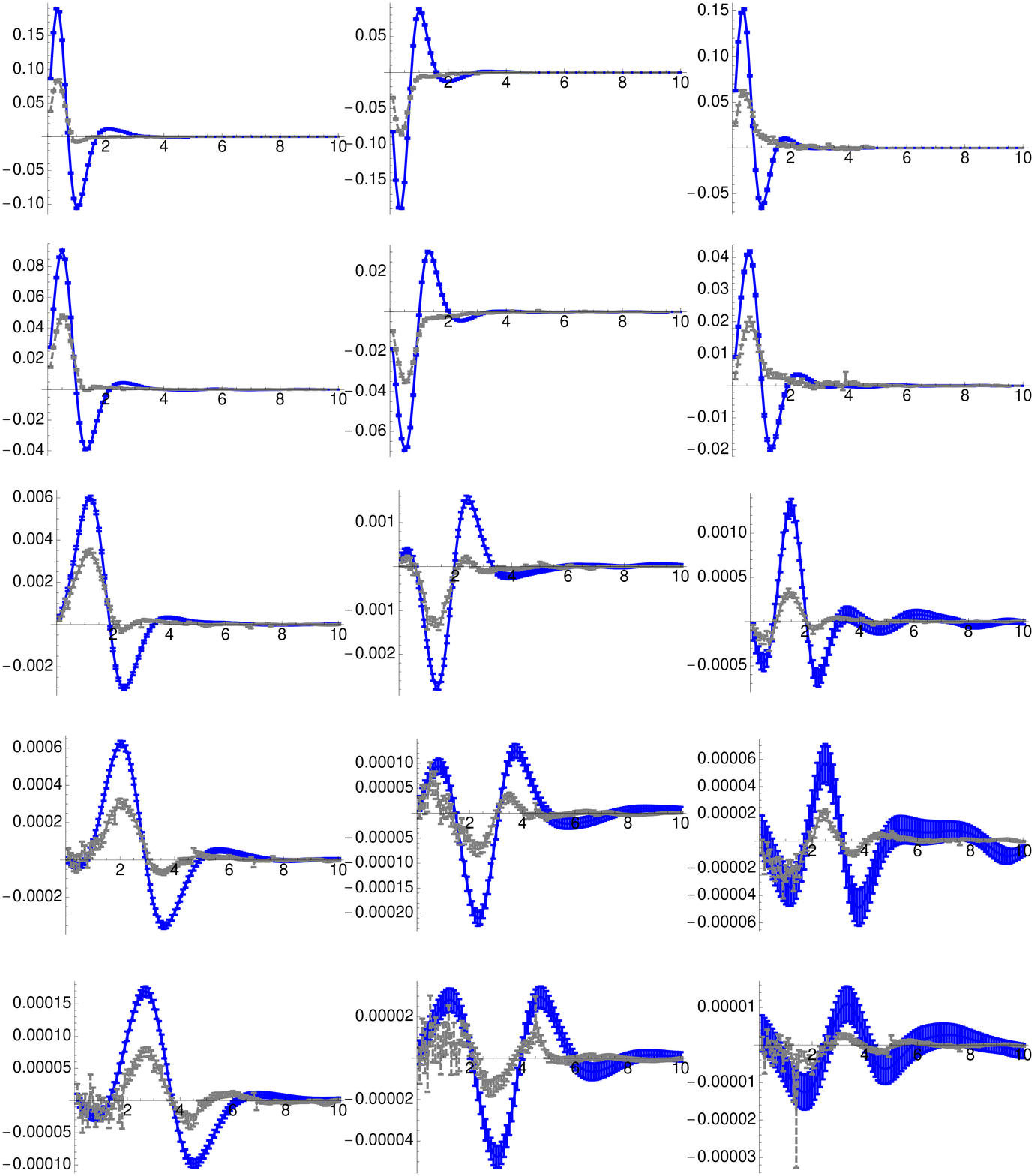,width=4.8in,clip=,angle=0}
\end{array}$
\caption{\label{plot:e100n456} 
The dN/dxdk GLV results for individual orders in opacity 4,5,6 for a 100 GeV massless quark jet. Shown is the result for the full result including the classical cascade, and the result only for the quantum source term. 
}
\end{figure*}

\begin{figure*}[htb] 
\centering
$\begin{array}{cc}
&
\epsfig{file=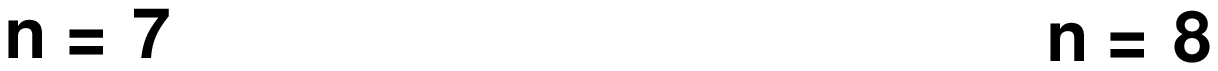,width=4.8in,clip=,angle=0} \\
\epsfig{file=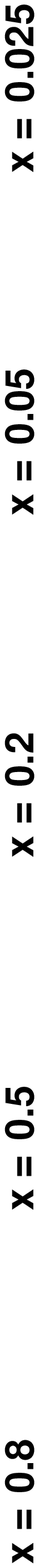,width=0.2in,clip=,angle=0} &
\epsfig{file=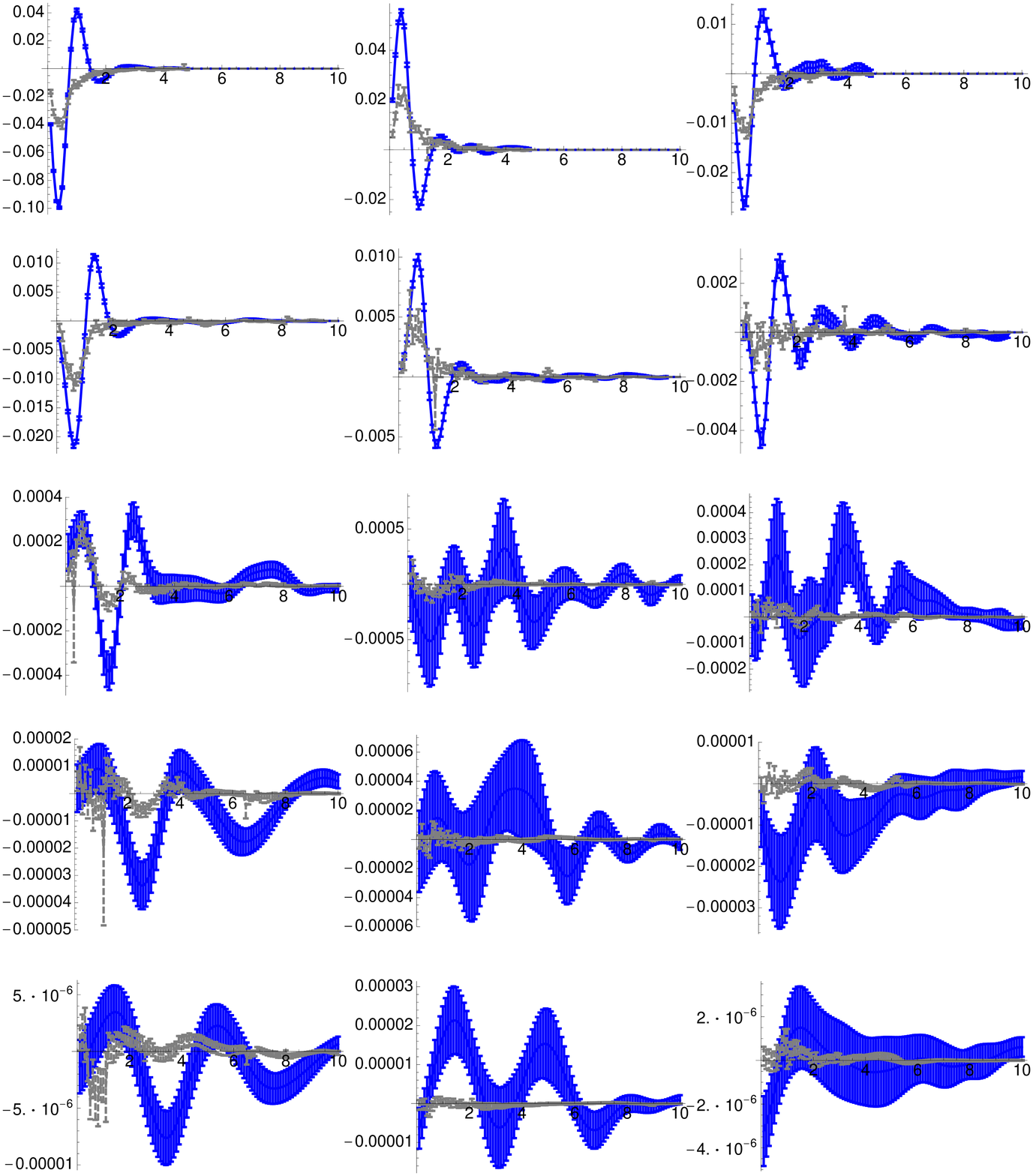,width=4.8in,clip=,angle=0}
\end{array}$
\caption{\label{plot:e100n789} 
The dN/dxdk GLV results for individual orders in opacity 7,8,9 for a 100 GeV massless quark jet. Shown is the result for the full result including the classical cascade, and the result only for the quantum source term. 
}
\end{figure*}

\clearpage
\section{dN/dxdk for p=100 GeV}
\begin{figure*}[htb!] 
\centering
$\begin{array}{c}
\epsfig{file=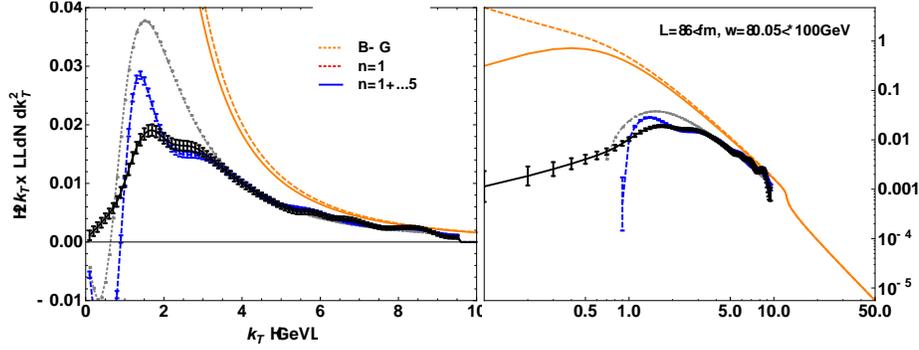,width=5in,clip=,angle=0}
\end{array}$
\caption{\label{plot:radNth1} 
The effect of the curtailment of the opacity series. The emitted gluon is at x=0.05, ie 5 GeV. The results are compared to the Bertsch-Gunion incoherent limits.
}
\end{figure*}

\begin{figure*}[htb!] 
\centering
$\begin{array}{c}
\epsfig{file=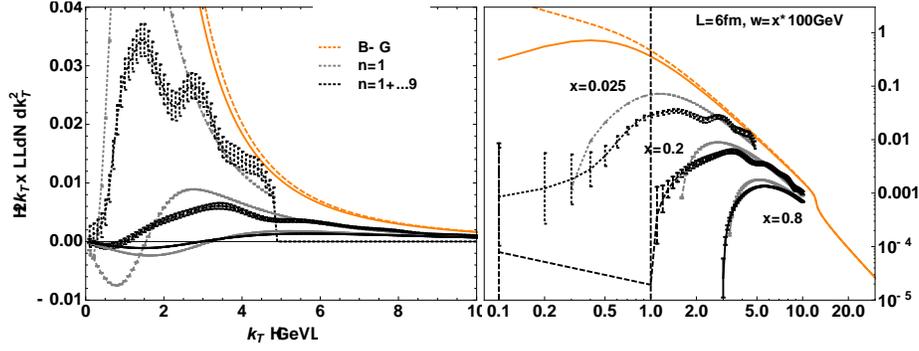,width=5in,clip=,angle=0}
\end{array}$
\caption{\label{plot:radNth2} 
The radiation spectrum $dN/dxdk_\perp$ summed to different orders for different values of $x$. The results are weighed by $x/L$, to give converging results for large $k_\perp$ - the Bertsch Gunion, incoherent answer scales with $L/x$.
}
\end{figure*}

\begin{figure*}[htb!] 
\centering
$\begin{array}{c}
\epsfig{file=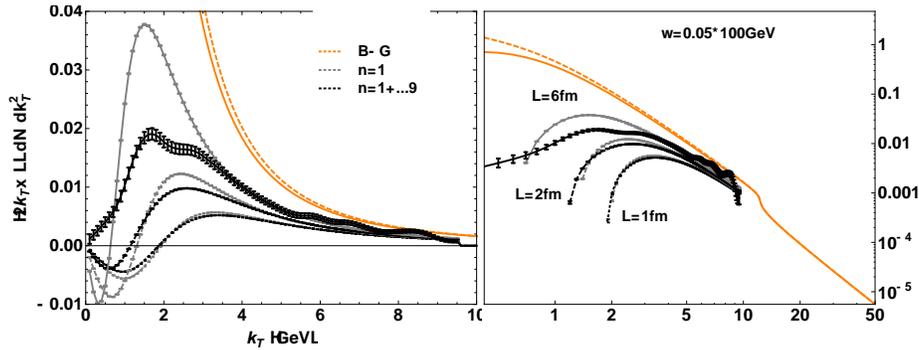,width=5in,clip=,angle=0}
\end{array}$
\caption{\label{plot:radNth3} 
The length dependence of the $dN/dxdk_\perp$ (at fixed x) radiation spectrum summed to different orders for a gluon with x=0.05 (ie 5 GeV). The answer is weighed by $x/L$, as the incoherent answer scales with $L/x$. The result increases with length, giving a length dependence stronger than $\propto L$, but the summed result also deviates further from the first order result.
}
\end{figure*}

\begin{figure*}[htb] 
\centering
$\begin{array}{c}
\epsfig{file=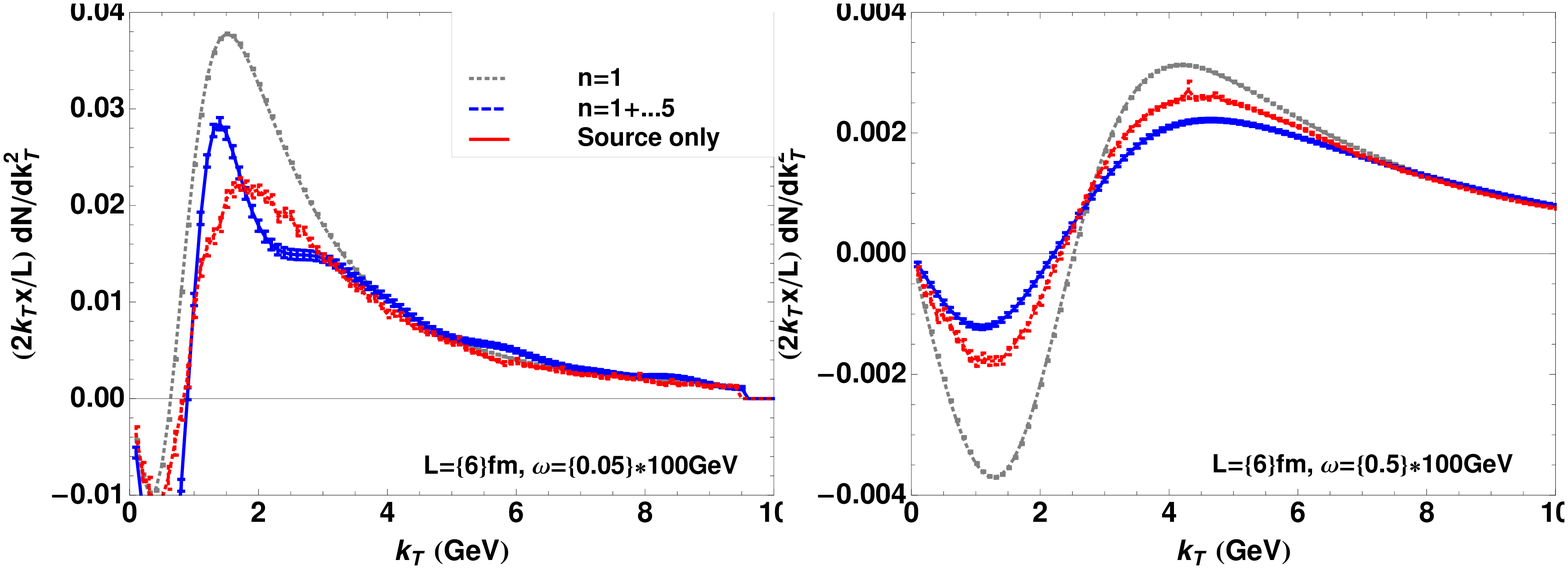,width=5in,clip=,angle=0} \\
\epsfig{file=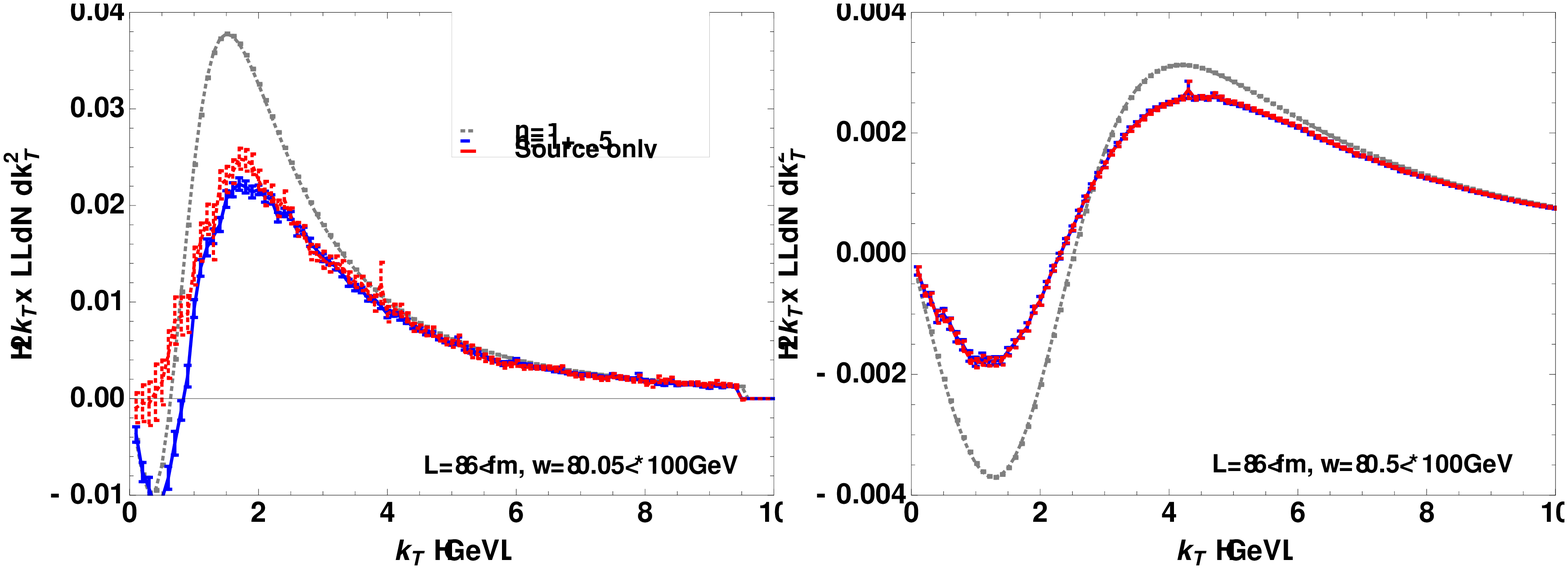,width=5in,clip=,angle=0}
\end{array}$
\caption{\label{plot:radNth4} 
The radiation spectrum $dN/dxdk_\perp$ at fixed x summed to different orders, with (\eq{eqn:allorders}) and without (\eq{eqn:allorders2}) the classical diffusion terms.
}
\end{figure*}

\clearpage
\section{dN/dxdk for p=20 GeV}

\begin{figure*}[htb] 
\centering
$\begin{array}{c}
\epsfig{file=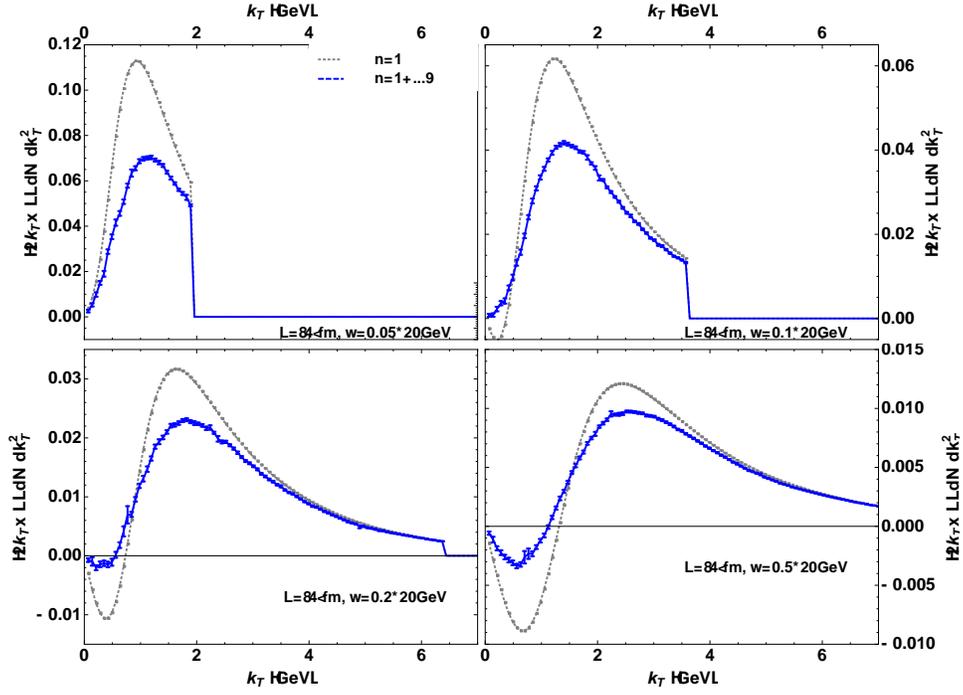,width=5in,clip=,angle=0}
\end{array}$
\caption{\label{plot:rade20} 
Radiative distributions (quantum source term only, \eq{eqn:allorders2}) for a massless quark jet of momentum 20 GeV. The sharp cut at small x is the kinematic cut on $k_\perp$, given by Djordjevic as $k_{max} = 2 x (1-x) E$.
}
\end{figure*}

\begin{figure*}[htb] 
\centering
$\begin{array}{c}
\epsfig{file=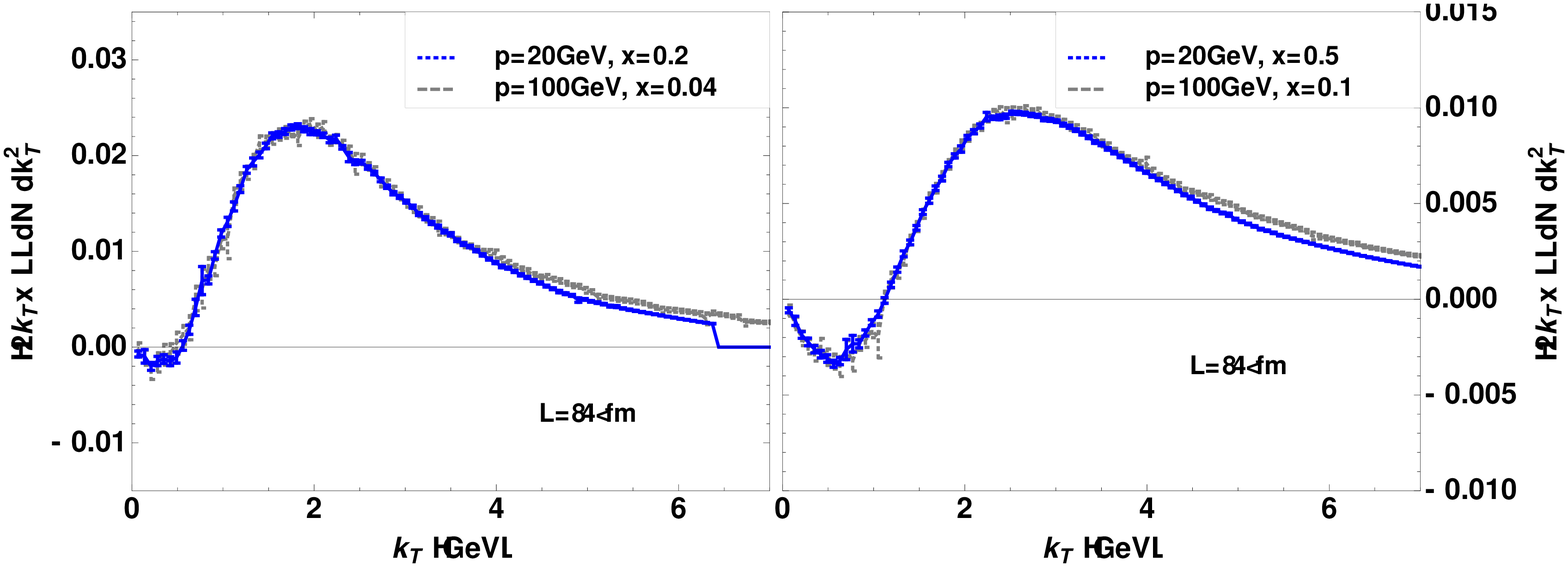,width=5in,clip=,angle=0}
\end{array}$
\caption{\label{plot:rade20vse100} 
The jet energy dependence of the emitted gluon distribution, for a fixed gluon energy with only the quantum source term, \eq{eqn:allorders2}. The distributions are very similar, except for a change in the $k_{max}$ and a small effect of changing $q_{max} = \sqrt{6 T E}$.
}
\end{figure*}

\clearpage
\section{GLV vs DGLV}

\begin{figure*}[htb] 
\centering
$\begin{array}{c}
\epsfig{file=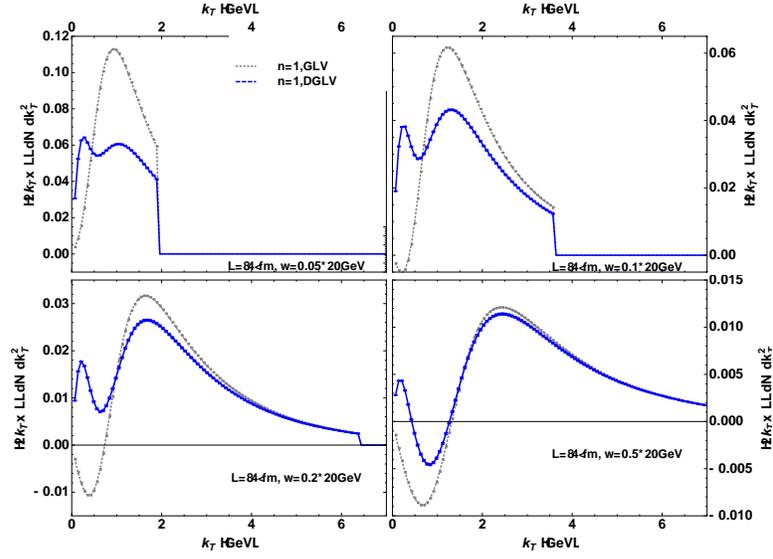,width=4in,clip=,angle=0}
\end{array}$
\caption{\label{plot:rade20dglvn1} 
First order in opacity radiative result, for GLV ($M=m_g=0$) and DGLV($M_{jet}=\mu/2$ = 0.25 GeV, $m_g = \mu/\sqrt{2}$=0.354 GeV), for the quantum source term only , \eq{eqn:allorders2}. The inclusion of the masses has a surprisingly large effect for gluon energies and $k_\perp$s up to $\approx 3$GeV.
}
\end{figure*}

\begin{figure*}[htb] 
\centering
$\begin{array}{c}
\epsfig{file=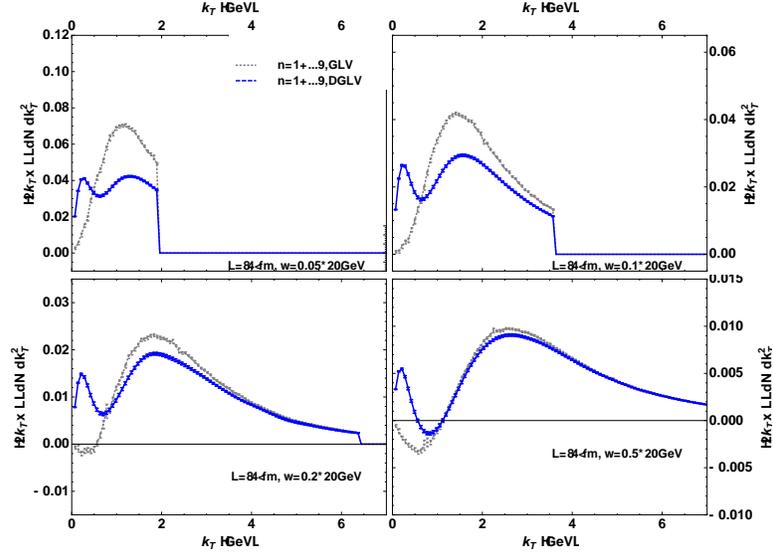,width=4in,clip=,angle=0}
\end{array}$
\caption{\label{plot:rade20dglvn9} 
Similar to the previous plot (GLV vs DGLV comparison \fig{plot:rade20dglvn1}), but for the result summed up to ninth order in opacity (again, quantum source term only, \eq{eqn:allorders2}).
}
\end{figure*}

\end{appendix}
\clearpage
\twocolumngrid

\end{document}